# Schedule generation schemes for the job-shop problem with sequence-dependent setup times: dominance properties and computational analysis

**Christian Artigues[1], Pierre Lopez[2] and Pierre-Dimitri Ayache[1]**

[1]*Laboratoire d'Informatique d'Avignon, FRE CNRS 2487,*

*BP 1228, 84911 Avignon, France*

*christian.artigues@lia.univ-avignon.fr*

*tel: +33 4 90 84 35 52 fax: +33 4 90 84 35 01*

[2]*LAAS - CNRS,*

7 avenue du Colonel Roche, 31077 Toulouse, France

lopez@laas.fr

**Abstract**

We consider the job-shop problem with sequence-dependent setup times. We focus on the formal definition of schedule generation schemes (SGSs) based on the semi-active, active, and non-delay schedule categories. We study dominance properties of the sets of schedules obtainable with each SGS. We show how the proposed SGSs can be used within single-pass and multi-pass priority rule based heuristics. We study several priority rules for the problem and provide a comparative computational analysis of the different SGSs on sets of instances taken from the literature. The proposed SGSs significantly improve previously best-known results on a set of hard benchmark instances.

## Introduction

The greedy constructive heuristics used to solve a NP-hard scheduling problem are mostly based on priority, also called dispatching, rules. A priority rule is used to select the task to be scheduled at each step of the algorithm, among the set of unscheduled tasks. Such heuristics are referred to as priority rule based algorithms. In some cases, the tasks are selected according to a predetermined order, which is equivalent to assign to each task a unique priority value. Choosing a priority rule, or a priority vector, is not sufficient to design a constructive scheduling heuristic. Indeed, the set of candidate tasks for being scheduled has to be defined, as well as the way the task selected based on a priority rule is scheduled. This logical process of the heuristic is called the schedule generation scheme (SGS), see Kolisch (1995) and Kolisch (1996)[13]. When dealing with a scheduling problem, designing SGSs is fundamental, not only for priority rule based heuristics. Indeed the SGS can also be viewed as the branching scheme of a branch and bound method. Branching consists in selecting a different task among the set of candidate tasks. It is of theoretical interest to study the dominance properties of the set of schedules that can be obtained by a given SGS, i.e. the nodes of the associated tree. Considering a scheduling objective, a set of schedules is said to be dominant when it contains at least one optimal schedule. Hence, the dominance property of the set of schedules obtainable with a given SGS gives the theoretical ability of the SGS to reach the optimum. By extension it can be stated that a SGS is dominant if there exists a priority vector leading to an optimal schedule when applied in conjunction with this SGS. It is well known that for scheduling problems with a regular objective function, i.e. non decreasing with task completion times, the sets of so-called semi-active and active schedules are dominant for any regular objective function whereas the set of so-called non-delay schedules is in general not dominant.

Several SGSs, and their corresponding schedule set, have been studied by Baker (1974) for the classical job-shop problem (JSP) where the tasks to be scheduled are called operations and are gathered into jobs. In this problem, the operations of a job are totally ordered so that no operation of a job can start before the completion of its predecessor. Furthermore, each operation requires a machine and each machine can process only one operation simultaneously. For the JSP, the Semi-Active SGS generates the dominant set of semi-active schedules, the SGS proposed by Giffler and Thompson (1960) and the serial SGS described in Kolisch (1995) generate the dominant set of active schedules, the Non-Delay SGS generates the non-dominant but smaller set of non-delay schedules.

It has been shown that the introduction of sequence-dependent setup times in a scheduling problem can call some dominance properties into question. Such a setup time may be needed between two operations sharing the same machine. Its duration depends on the preceding and following operations. For parallel machine problems, Ovacik and Uzsoy (1993) have proved that the set of non-delay schedules, dominant for this particular problem, is not dominant anymore in the presence of setup times. This has brought Schutten (1996) to provide for the same problem an alternative list scheduling algorithm, able to generate a set of dominant schedules. Recently, Hurink and Knust (2001) stated that a dominant list scheduling algorithm is even unlikely to exist for the parallel machine problem involving sequence-dependent setup times and precedence relations.

The job-shop problem with sequence-dependent setup times (SDST-JSP) has been studied by a few authors. Schutten (1998) proposes an extension of the disjunctive graph model and the shifting bottleneck procedure. Branch and bound algorithms are proposed by Gupta (1986) and by Brucker and Thiele (1996). Focacci, Laborie and Nuijten (2000) propose a Constraint Programming based method. An insertion heuristic is proposed by Sotskov, Tautenhahn and Werner (1999) for the job-shop problem with sequence-independent batch setup times and by Artigues and Roubellat (2002) for the SDST-JSP. Priority rules have been tested by using the non-delay schedule generation scheme by Kim and Bobrowski (1994) and Noivo and Ramalhinho-Lourenço (1998) while Ovacik and Uzsoy (1994) propose a priority rule based heuristic aiming at exploiting real time shop floor status information, working in the set of active schedules. Brucker and Thiele (1996) propose several priority-rule based heuristics using a semi-active SGS and an extended Giffler-Thompson SGS. For more information on scheduling research involving setups, we refer to the surveys of Allahverdi, Gupta and Aldowaisan (1999) and Yang and Liao (1999). None of these studies is concerned with the definition or the evaluation of schedule generation schemes for the SDST-JSP. In this paper, we propose to fill this gap by providing in a formal way several SGSs for the SDST-JSP. We study the dominance properties of the corresponding schedule sets and we give computational results comparing the different SGSs used as single- and multi-pass priority rule based heuristics using various priority rules to results obtained by exact methods.

The paper is organized as follows. In Section 1, we define the SDST-JSP. In Section 2, the concepts of non-delay, active, and semi-active schedules are extended; several SGSs and their dominance properties are given. Computational experiments are discussed in Section 3. Concluding remarks are drawn in Section 4.

## 1. The job-shop problem with sequence-dependent setup times

We define in this Section the job-shop problem with sequence-dependent setup times (SDST-JSP). As in the classical job-shop problem, a set $J$ of $n$ jobs $1,\ldots, n$ has to be performed by a set $M$ of $m$ machines $1,\ldots, m$. Each job $i$ is made of $m$ operations $o(i,1),\ldots, o(i,m)$. $N = mn$ denotes the total number of operations. The decision variables of this problem are the operation starting times $t_{o(i,j)}$, $\forall i=1,\ldots, n$; $\forall j=1,\ldots, m$.

Each operation $o(i,j)$ has to be scheduled on a machine $m_{o(i,j)}$ during $p_{o(i,j)} \geq 0$ uninterrupted time units such that any machine can process only one operation at the same time. For notation convenience, we also assume that:

- Operations are indexed from 1 to $N$, i.e. we have $o(i,j) = j+(i-1)m$; it follows that we can refer to an operation by its own index $o\in\{1,..., N\}$. $m_o$, $p_o$, $t_o$ denote the machine, duration and start time of operation $o$, respectively. $O=\{1,\ldots, N\}$ denotes the set of operations.
- An initial dummy operation 0 is defined as the first operation of each job: $t_0 = 0$; $p_0 = 0$. We assume that $o(i,0) = 0$, $\forall i=1,\ldots, n$.

Switching from an operation $a\in O$ to an operation $b\in O$ on machine $m_a = m_b$, requires a non-negative setup time $s_{a,b}$. Moreover, an initial non-negative setup time $s_{0,a}$ is required if operation $a$ is the first to be sequenced on machine $m_a$.

The problem can be formulated as follows:

(1) $\quad$ Min $C_{\max}$

Subject to:

(2) $\quad C_{\max} \geq t_{o(i,m)} + p_{o(i,m)} \qquad (i = 1,\ldots, n)$

(3) $\quad t_o \geq s_{0,o} \qquad (o \in O)$

(4) $\quad t_{o(i,j)} \geq t_{o(i,j-1)} + p_{o(i,j-1)} \qquad (i = 1,\ldots, n; j = 2,\ldots, m)$

(5) $\quad t_a \geq t_b + p_b + s_{b,a}$ or $t_b \geq t_a + p_a + s_{a,b} \qquad (a,b \in O, a \neq b, m_a = m_b)$

A feasible schedule is defined by a vector of starting times $t = (t_o)_{o\in O}$ compatible with constraints from (3) to (5). Constraints (3) ensure that the initial setup time is satisfied on each machine. Constraints (4) ensure that precedence constraints within each job are satisfied. Constraints (5) are the disjunctive constraints that prevent two operations using the same machine from being scheduled simultaneously, the necessary setup times between the two operations being respected. Note that constraints (4) allow setup anticipation, i.e. the start of the setup for an operation on the machine while the preceding operation in the job is not completed. The objective function (1) is to minimize the makespan $C_{\max}$ defined by constraints (2). In addition, it is assumed that setup times verify the triangle inequality:

(6) $$s_{a,b} + s_{b,c} \geq s_{a,c} \qquad (a,b,c \in O,\ a \neq b \neq c,\ m_a = m_b = m_c)$$

This common assumption is reasonable according to practical applications (e.g. Brucker and Thiele (1996)). It is easy to show that the SDST-JSP is NP-hard. Indeed, setting all setup times to 0 gives the classical JSP. Setting $m = 1$ and $p_o = 0$, $\forall o \in O$ gives the travelling salesman problem.

*Example* 1

Let us consider the following example with 2 machines and 2 jobs. Setup times are present on machine 1 whereas there is no setup time needed for machine 2. The durations are $p_{o(1,1)} = 2$, $p_{o(1,2)} = 1$, $p_{o(2,1)} = 5$, $p_{o(2,2)} = 2$. The machine requirements of the jobs are $m_{o(1,1)} = 1$, $m_{o(1,2)} = 2$, $m_{o(2,1)} = 2$ and $m_{o(2,2)} = 1$. The setup times on machine 1 are $s_{0,o(1,1)} = 1$, $s_{0,o(2,2)} = 2$, $s_{o(1,1),o(2,2)} = 10$, $s_{o(2,2),o(1,1)} = 3$. The triangle inequality holds.

Supprimé : are

A feasible solution to example 1 problem of makespan $C_{\max} = 15$ is depicted in figure 1. The required setup times are displayed in grey. Setup anticipation is illustrated on machine 1 where the setup for $o(2,2)$ starts while $o(2,1)$ is not even started.

***** INSERT FIGURE 1 ABOUT HERE *****

## 2. Schedule generation schemes for the SDST-JSP

### 2.1. General principles and definitions

In this Section, we present the framework of all Schedule Generation Schemes considered in this paper.

*Greedy construction process*

The category of schedule generation schemes we consider builds a solution to the scheduling problem in exactly $N$ steps. At each step, exactly one operation is transferred from the set of unscheduled operations, denoted by $Q$, to the set of scheduled operations. At the beginning we have $Q = O$ and the process stops when $Q$ becomes empty.

At the step where operation $o(i,j)$ is scheduled, i.e. removed from $Q$, its start time $t_{o(i,j)}$ is assigned to a value that will not be modified during the remaining of the construction process (greedy behaviour).

All SGSs use at each of the $N$ steps a set of *eligible* operations $E$, called the eligible set, which is a subset of $Q$ restricting the choice for the operation to be scheduled. More precisely, the eligible set $E$ is a subset of the *available* set $A$ defined as follows. An unscheduled operation belongs to the available set either if it is the first operation of its job or if its preceding operation has been scheduled in a previous step. An operation having this property is called available. Hence we have $E \subseteq A \subseteq Q$. The SGSs under study differ in the additional restrictions they use to define the eligible set $E$.

Last, if the SGS is used as a branching scheme of a branch and bound algorithm, branching consists in selecting an operation in $E$ as the next operation to be scheduled. Hence, the number of children of a given node is $|E|$.

The general framework of the considered SGSs given a priority vector $\pi$ is summarized in figure 2. Steps 4 and 5 correspond to the priority rule computation for the eligible operations and the earliest feasible start time computation of the selected operation, respectively. These concepts are explained thereafter. We have to underline that this framework is restrictive since other SGSs could be defined where a non available operation could be selected for scheduling, or/and where the start time of the scheduled operation could be modified at a later step. Nevertheless the well-known semi-active, non-delay and active SGSs for the classical job-shop problem (Baker (1974)), the serial and parallel SGSs [13]) and the strict order SGS proposed by Carlier and Néron (2000) for the resource-constrained project scheduling problem all belong to the considered category.

******************** INSERT FIGURE 2 HERE ************************

*Computation of earliest feasible start times: appending vs. insertion SGSs*

The operation selected in the eligible set $E$ is scheduled at its earliest feasible start time at step 5 of the general framework presented in figure 2. For the computation of these start times, we make a distinction between *appending* schedule generation scheme and *insertion* schedule generation scheme. In an appending scheme, an unscheduled operation can be scheduled only after all the operations already scheduled on its machine, without delaying these operations. In an insertion scheme, the unscheduled operation may be inserted somewhere in the sequence of operations already scheduled on its machine. In a SGS it is necessary to compute the earliest resource- and precedence-feasible start time of unscheduled operations. We give hereafter two algorithms for computing the start time of an eligible operation $o(i,j)\in E$ depending on the SGS category.

Let $k = m_{o(i,j)}$ and let $\lambda(k) \in \{1,\ldots, N\}$ denote the latest operation (at the current iteration) scheduled on machine $k$. When no operation has been scheduled yet on machine $k$, $\lambda(k)$ is set to the dummy first operation 0. Let $c_{\lambda(k)}$ be the completion time of operation $\lambda(k)$. For each scheduled operation $o(x,y)\in O\backslash Q$, let $c_{o(x,y)} = t_{o(x,y)}+p_{o(x,y)}$ denote its completion time. Let $\mathrm{ESA}_{o(i,j)}$ denote the earliest possible start time of an unscheduled operation $o(i,j)\in Q$ in an appending schedule generation scheme. $\mathrm{ESA}_{o(i,j)}$ can be computed in $O(1)$ at any step of the constructive process by:

$$\mathrm{ESA}_{o(i,j)} = \max\left( c_{o(i,j\text{-}1)}, c_{\lambda(k)}+s_{\lambda(k),o(i,j)}\right) \text{ with } k = m_{o(i,j)} \tag{7}$$

Note that (7) can be easily derived from (4) and (5) under the triangle inequality assumption (6).

Let $\eta_k$ denote the number of operations scheduled on machine $k$. Let $\sigma$ denote the partial sequence made of operations $\sigma(0,k) = 0$, $\sigma(1,k)$, …, $\sigma(\eta_k,k) = \lambda(k)$, scheduled on machine $k$. For an insertion SGS, since we consider that the start time of a scheduled operation cannot be changed, we search for the first insertion position where the selected operation can fit without delaying the subsequent operation. Formally, let $\mathrm{ESI}_{o(i,j)}$ denote the earliest possible start time of an unscheduled operation in an insertion schedule generation scheme.

A position $q$ with $0 \le q \le \eta_k$ verifying

$$\max\left( c_{o(i,j\text{-}1)}, c_{\sigma(q,k)}+s_{\sigma(q,k),o(i,j)}\right) + p_{o(i,j)} + s_{o(i,j),\sigma(q+1,k)} \le t_{\sigma(q+1,k)} \tag{8}$$

is a feasible insertion position between operation $\sigma(q,k)$ (possibly dummy first operation 0) and operation $\sigma(q+1,k)$. Again, (8) can be easily derived from (4) and (5) under the triangle inequality assumption (6). If there exists at least one position $q$ verifying equation (8) we set $\tilde{q} = \min_{q \text{ verifies } (8)} q$ and we have:

$$\mathrm{ESI}_{o(i,j)} = \max\left( c_{o(i,j\text{-}1)}, c_{\sigma(\tilde{q},k)}+s_{\sigma(\tilde{q},k),o(i,j)}\right) \tag{9}$$

Otherwise we set

$$\mathrm{ESI}_{o(i,j)} = \mathrm{ESA}_{o(i,j)} \tag{10}$$

Obviously, the earliest start time of an eligible operation in an insertion scheme can be computed in $O(m)$ since there are at most $m$-1 operations scheduled on machine $k=m_{o(i,j)}$.

*Priority rules*

If the SGS is used as a simple contructive algorithm, we suppose that each operation $o(i,j)$ of $E$ has a priority value $\pi_{o(i,j)}$. This priority value can be computed in a static way (it remains constant throughout the process) or in a dynamic way (it may have to be recomputed at each step of the SGS). We assume without loss of generality that, at each iteration, the SGS selects the operation with the smallest priority value as stated at step 4 of the general framework displayed in figure 2. In the case where the priority is static, the operations can be pre-sorted in a non-decreasing order of their priority value and form a list. Since such predefined lists and static priority rules are strictly equivalent, we restrict this study to the priority rules.

*Schedule category*

The way the set $E$ is computed by a given SGS is mostly driven by the restriction of the search to a specific category of schedule. Let $F$ denote the set of solutions of the SDST-JSP satisfying constraints (3), (4) and (5), i.e. the set of feasible solutions. $F$, which is the set of vectors $t$ verifying the problem constraints, possibly represents a huge search space. Hence, many works on scheduling theory rely on the restriction of the search on smaller subsets of $F$, which defines categories of schedules. Among these categories, most of the existing approaches focus on the sets of *semi-active*, *non-delay*, and *active* schedules. Such concepts have been introduced for the job-shop problem without setup times by Baker (1974). More recently, they have been extended to the resource-constrained project scheduling problem by Sprecher, Kolisch and Drexl (1995). By extension, non-delay, semi-active and active SGSs generate only non-delay, semi-active and active schedules, respectively. A set of schedules of a given category is said to be dominant w.r.t. makespan minimization if it contains at least one optimal solution. In the case where the SGS is used as a branching scheme of an implicit enumeration tree, the SGS may have the property to generate all the schedules of its corresponding category. By extension, a dominant SGS corresponds to a dominant category of schedules and has the latter property. In the remaining, we discuss the extension of the schedule categories and the corresponding schedule generation scheme for the SDST-JSP.

## 2.2. Semi-active SGSs

Following Sprecher, Kolisch and Drexl (1995), let us define a local left-shift of an operation $o(i,j)$ in a feasible schedule $t$ as the move giving schedule $t'$ where $t'_{o(i,j)} = t_{o(i,j)} - 1$ and $t'_{o(x,y)} = t_{o(x,y)}$ $\forall$ $o(x,y) \in O \backslash \{o(i,j)\}$.

*Definition 1*

A semi-active schedule is a feasible schedule in which no local left-shift leads to another feasible schedule.

It follows that in a semi-active schedule, any operation has at most one tight job- or resource-precedence constraint, preventing it from being locally left-shifted. From any feasible non semi-active schedule, there always exists a (possibly empty) series of local left-shifts giving a semi-active schedule without increasing the makespan. Hence, the set of semi-active schedules is smaller than the set of feasible schedules and is dominant for the SDST-JSP.

The semi-active appending schedule generation schemes aim at generating schedules belonging to the set of semi-active schedules. The SGS able to generate all semi-active schedules is noted SemiActiveSGS. In the general framework defined in Section 2.1, it has the following particularities:

- SemiActiveSGS is an appending SGS;
- the eligible set $E$ is equal to the entire available set $A$.

Basically, the semi-active SGS selects an available operation and schedules it as soon as possible on its machine, after the previously scheduled operation. The overall time complexity of the semi-active schedule is $O(NnT)$, $T$ being the time complexity of the priority rule computation.

*Lemma* 1

For any priority vector, SemiActiveSGS generates a semi-active schedule and any semi-active schedule is generated by SemiActiveSGS with a priority vector.

*Proof*

We first demonstrate that for a given priority vector $\pi$, schedule $t$ generated by SemiActiveSGS is a semi-active schedule. Suppose that $t$ is not semi-active. Hence, there exists $o(i,j) \in O$ such that $\tau$ is a feasible schedule where $\tau_{o(i,j)} = t_{o(i,j)} - 1$ and $\tau_{o(x,y)} = t_{o(x,y)}$, $\forall$ $o(x,y) \in O \backslash \{o(i,j)\}$. Since $t_{o(i,j)}$ =ESA$_{o(i,j)}$, we have $t_{o(i,j)} = \max\ (c_{o(i,j)},\ c_a + s_{a,o(i,j)})$ where $a$ is the operation sequenced right before $o(i,j)$ on its machine (which is valid according to the triangle inequality assumption). Hence setting $\tau_{o(i,j)} = t_{o(i,j)} - 1$ is unfeasible and $t$ is semi-active. Conversely, let $t$ be a feasible semi-active schedule. Sort the operations in the order of non-decreasing start times and set $\pi_{o(i,j)}$ to the rank of $o(i,j)$. Obviously if $\eta_k$ is the number of operations scheduled on machine $k$, we have $\pi_{\sigma(1,k)} < \pi_{\sigma(2,k)} < \ldots < \pi_{\sigma(\eta_k,k)}$ and the schedule obtained by SemiActiveSGS is $t$.

□

*Theorem* 1

The set of schedules generated by SemiActiveSGS is dominant.

*Proof*

From lemma 1, SemiActiveSGS is able to generate all semi-active schedules. Furthermore, the set of semi-active schedules is dominant.

□

## 2.3. Active SGSs

Sprecher, Kolisch and Drexl (1995) define a global left-shift of an operation $o(i,j)$ in a feasible schedule $S$ as a move giving schedule $S'$ after performing one or several local left-shifts of $o(i,j)$.

*Definition 2*

An active schedule is a feasible (semi-active) schedule in which no global left-shift leads to a feasible schedule.

It follows from Definition 2 that in an active schedule, for any operation $o(i,j)$, one cannot find any feasible insertion position, located before $o(i,j)$ in the sequence of operations scheduled on $m_{o(i,j)}$, large enough to receive $o(i,j)$. From any semi-active non active schedule, one can always obtain an active schedule by performing a series of global left-shifts. Hence, the set of active schedules is smaller than the set of semi-active schedules and remains dominant for the SDST-JSP.

To provide an active SGS, it seems natural to study the extension of the Giffler-Thompson Algorithm (Giffler and Thompson (1960)) which is the most famous active schedule generation scheme for the classical job-shop problem. Before looking at possible extensions, let us recall the behaviour of this well-known algorithm in the case of a job-shop problem with zero setup times. The Giffler-Thompson algorithm (GifflerThompsonSGS) has the following particularities in the general framework defined in Figure 2:

- GifflerThompsonSGS is an appending SGS
- the eligible set $E$ is computed as follows. Let $\tilde{o}$ denote the available operation of smallest completion time (job index is used to break ties). We have:

(11) $$\tilde{o} = \text{argmin} \{\text{ESA}_o + p_o \mid o \in A\}$$

$E$ is the set of operations conflicting with $\tilde{o}$ that is

(12) $$E = \{o \mid o \in A,\ m_o = m_{\tilde{o}},\ \text{ESA}_{\tilde{o}} + p_{\tilde{o}} > \text{ESA}_o\}$$

In the case of the job-shop with zero setup times, any schedule generated by GifflerThompsonSGS is active because no unscheduled operation $o$ can be inserted before the selected operation $o^*$. Indeed, we have $\text{ESA}_{\tilde{o}} + p_{\tilde{o}} > \text{ESA}_{o^*}$ since $o^*$ belongs to the eligible set defined above and also $\text{ESA}_o + p_o \geq \text{ESA}_{\tilde{o}} + p_{\tilde{o}}$ because $o$ is unscheduled. Hence we have $\text{ESA}_o + p_o > \text{ESA}_{o^*}$, which makes the insertion of $o$ before $o^*$ impossible. Furthermore, it is well-known that using GifflerThompsonSGS as a branching scheme (branching on operations of $E$) gives all active schedules.

We show hereafter that extending the Giffler-Thompson algorithm to non-zero setup times raises significant problems. Brucker and Thiele (1996) propose for the SDST-JSP a SGS directly inspired from the Giffler-Thompson algorithm by using (12) as conflict set definition, which lies in ignoring the possible setup times. Taking account of setup times, the set $E$ can be defined as

(13) $$E = \{o \mid o \in A,\ m_o = m_{\tilde{o}},\ \text{ESA}_{\tilde{o}} + p_{\tilde{o}} + s_{\tilde{o},o} > \text{ESA}_o\}$$

i.e. the set of available operations whose earliest appending start time has to be increased if $\tilde{o}$ is scheduled. Unfortunately, coupling this definition of set $E$ with an appending SGS does not necessarily lead to an active schedule. Indeed, suppose that an operation $o^* \neq \tilde{o}$ is selected for scheduling. Nothing prevents another available operation $o$ to be insertable before $o^*$ although conflicting with $\tilde{o}$ provided that it verifies $\text{ESA}_o + p_o + s_{o,o*} \leq \text{ESA}_{o*}$. This is illustrated by the following second example.

*Example* 2

We consider an example problem with 2 machines and 4 jobs. Setup times are present on machine 1 whereas there is no setup time needed for machine 2. The considered durations are $p_{o(1,1)} = 2$, $p_{o(2,1)} = 5$, $p_{o(2,2)} = 2$, $p_{o(3,1)} = 2$, $p_{o(4,1)} = 3$. Other operation durations are not necessary for the present illustration and are left undetermined. The machine requirements of the jobs are $m_{o(1,1)} = 1$, $m_{o(1,2)} = 2$, $m_{o(2,1)} = 2$, $m_{o(2,2)} = 1$, $m_{o(3,1)} = 2$, $m_{o(3,2)} = 1$, $m_{o(4,1)} = 1$, $m_{o(4,2)} = 2$. The setup times on machine 1 are $s_{0,o(1,1)} = 1$, $s_{0,o(2,2)} = 2$, $s_{0,o(4,1)} = 1$, $s_{o(1,1),o(2,2)} = 10$, $s_{o(1,1),o(4,1)} = 9$, $s_{o(2,2),o(1,1)} = 8$, $s_{o(2,2),o(4,1)} = 5$, $s_{o(4,1),o(1,1)} = 3$, $s_{o(4,1),o(2,2)} = 1$. Setup times involving $o(3,2)$ are not necessary for the present illustration. The triangle inequality holds.

Suppose that operation $o(2,1)$ is selected for scheduling at the first step of the procedure. This is compatible with the Giffler-Thompson algorithm since $o(2,1)$ is conflicting with $o(3,1)$, which has the smallest earliest completion time. Now at the second step of the procedure, available operations are $o(1,1)$, $o(2,2)$, $o(3,1)$ and $o(4,1)$. Among them, the operation having the smallest earliest appending completion time, computed by Equation 12, is $\tilde{o}=o(1,1)$ with $\text{ESA}_{o(1,1)} + p_{o(1,1)} = 3$. Figure 3 illustrates the conflict definition at the considered step. Scheduled operation $o(2,1)$ is displayed in bold whereas available operations are displayed in plain line. Figure 3 displays 5 partial solutions (a), (b), (c), (d) and (e) all with $o(2,1)$ scheduled first on machine 2.

(a) $o(2,2)$ is scheduled first right before $o(1,1)$.
(b) $o(1,1)$ is scheduled first right before $o(2,2)$.
(c) $o(4,1)$ is scheduled first right before $o(1,1)$.
(d) $o(1,1)$ is scheduled first right before $o(4,1)$.
(e) $o(4,1)$ is scheduled first right before $o(2,2)$

********************* INSERT FIGURE 3 HERE ************************

Partial solutions (a), (b), (c) and (d) illustrate that, according to expression (13), the conflict set at the considered step is $E=\{o(1,1), o(2,2), o(4,1)\}$. Indeed it is clear that scheduling first $o(1,1)$ on machine 1 increases the earliest appending start time of $o(2,2)$, as shown by comparing (a) and (b), and also the earliest appending start time of $o(4,1)$ as shown by comparing (c) and (d). Suppose now that the priority rule selects $o(4,1)$ for scheduling, right before $o(2,2)$ as displayed in partial solution (e). It appears here that configuration (a) is not active because $o(2,2)$ has the same start time (5) if it is scheduled first or right after $o(4,1)$. The explanation is that $o(4,1)$ is not in conflict with $o(2,2)$ although both operations are in conflict with $o(1,1)$. In other words, the conflict relation is not transitive anymore in the presence of non zero setup times.

Hence defining an appending SGS with an eligible set computed by expression (13) does not necessary lead to active schedules. To overcome this drawback, we may use a more restrictive definition of the eligible set, ensuring that an eligible operation is in conflict with all other ones. Let us define the eligible set as the set of available operations scheduled on machine $m_{\tilde{o}}$ in conflict with another available operation:

(14) $$E = \{o \mid o \in A, \forall o' \in A, m_o = m_{o'} = m_{\tilde{o}}, \ \text{ESA}_{o'} + p_{o'} + s_{o',o} > \text{ESA}_o \}$$

With this definition, at the considered step of the example 2 problem (figure 3), the conflict set is reduced to $E = \{o(1,1), o(4,1)\}$ and the two corresponding partial schedules lead to active schedules. Again, a counterexample shows that, even with this more sophisticated conflict set definition, the SGS may generate a non active schedule.

*Example* 3

Let us modify example 2 by adding a 3rd machine with zero setup times and some other modifications. We suppose that jobs 1, 2 and 3 are unchanged except that $o(1,3)$, $o(2,3)$ and $o(3,3)$ require machine 3. Job 4 is now such that $m_{o(4,1)}=3$, $m_{o(4,2)}=1$ and $m_{o(4,3)}=2$ with $p_{o(4,1)}=3$, $p_{o(4,2)}=1$. $o(4,1)$ is no more subject to setup times whereas $o(4,2)$ has the following setup times : $s_{0,o(4,2)}=0$, $s_{o(1,1),o(4,2)}=9$, $s_{o(2,2),o(4,2)}=5$, $s_{o(4,2),o(1,1)}=3$ and $s_{o(4,2),o(2,2)}=1$. The remaining data is left unchanged or undetermined. Since setup times on machine 1 are the same as in example 2, the triangle inequality still holds. □

We again suppose that the priority rule selects first operation $o(2,1)$ for scheduling. Again this is in accordance with the Giffler-Thompson algorithm since $o(2,1)$ is in conflict with $o(3,1)$ which has the smallest earliest completion time at step 1. Now at step 2, available operations are $o(1,1)$, $o(2,2)$, $o(3,1)$ and $o(4,1)$. The operations with the smallest earliest appending completion time are $o(1,1)$ and $o(4,1)$. Using job index to break the tie we have $\tilde{o}= o(1,1)$. Operation $o(4,2)$ being unavailable, the conflict set according to definition 14 is $E$ =

$\{o(1,1),o(2,2)\}$. Figure 4 illustrates the conflict definition at the considered step. The scheduled operation $o(2,1)$ is displayed in bold. Available operations are displayed in plain line whereas non-available operation $o(4,2)$ is displayed in dotted line. The two corresponding partial solutions illustrating the conflict are (a) and (b). Suppose that at step 2 of the SGS, $o(2,2)$ is selected for scheduling. Suppose now that at step 3 of the SGS, operation $o(4,1)$ is selected for scheduling. Then, $o(4,2)$ becomes available at time 3 and can be inserted before $o(2,2)$ as shown in part (c). Partial solution (a) does not lead to an active solution if an appending scheme is used.

********************* INSERT FIGURE 4 HERE *************************

This counterexample of figure 4 shows that to obtain an active schedule with an appending SGS, the conflict set cannot be restricted to the available operations. Including non-available operations in the conflict set is out of the scope of the SGSs presented in this paper. However, another way of obtaining active schedules is to use an insertion SGS. This yields the following two extended Giffler-Thompson SGSs.

- GifflerThompsonSGS1 is an insertion SGS
- The conflict set is defined by expression (13).

or

- GifflerThompsonSGS2 is an insertion SGS
- The conflict set is defined by expression (14).

Note that when all setup times are equal to 0, the two proposed algorithms behave strictly like the Giffler-Thompson algorithm.

We introduce also another SGS, called the serial schedule generation scheme, [12] Kolisch (1995) and Kolisch (1996), which has been previously defined for the resource-constrained project scheduling problem. For the classical JSP, it is also able to generate all active schedules. Its extension to the SDST-JSP is straightforward.

- SerialSGS is an insertion SGS
- the eligible set $E$ is equal to the entire available set $A$.

The algorithm behaves like the semi-active SGS except that the selected operation is inserted at the earliest feasible position on its machine.

$T$ being the time complexity of the priority rule computation, the time complexity of ExtendedGifflerThompsonSGS1 and SerialSGS is $O(N(nT+m))$, while the time complexity of ExtendedGifflerThompsonSGS2 is $O(N(n^2T+m))$.

We now focus on the theoretical ability of the three active SGSs to generate all active schedules, i.e. their dominance properties. First, theorem 2 shows that all three proposed active SGSs generate active schedules.

*Theorem* 2

For any priority vector, any insertion SGS generates an active schedule.

*Proof*

Suppose that a schedule $t$ generated by an insertion SGS is not active. Let $\sigma$ denote the sequence made of operations $\sigma(0,k) = 0$, $\sigma(1,k)$, …, $\sigma(n,k)$, scheduled on each machine $k$ for schedule $t$. If $t$ is not active, then there must exist a machine $k$ and an operation $o(i,j)$ scheduled on $k$ at position $\rho\in\{2,\ldots,m\}$ such that expression (8) is verified for insertion position $q<\rho-1$, i.e. we have:

$\max\,(\,c_{o(i,j\text{-}1)}\,,\,c_{\sigma\,(q,k)}+s_{\sigma\,(q,k),o(i,j)})+p_{o(i,j)}+s_{\,o(i,j),\,\sigma(q+1,k)}\leq t_{\sigma\,(q+1,k)}$

with (a) $t'_{o(i,j)}=\max\,(\,c_{o(i,j\text{-}1)}\,,\,c_{\sigma(q,k)}+s_{\sigma(q,k),o(i,j)})<t_{o(i,j)}$.

Let $g$ denote the step where $o(i,j)$ is scheduled by the SGS. Let $\eta^g_k$ denote the number of operations scheduled on machine $k$ at step $g$. Let $\sigma^g$ denote the partial sequence made of operations $\sigma^g(0,k) = 0$, $\sigma^g(1,k)$, …, $\sigma^g(\eta^g_k,k)$ scheduled on machine $k$ at step $g$. For each of the possible cases concerning the components of $\sigma^g$, a contradiction is raised with (a).

- If $\sigma(q,k)$ is in $\sigma^g$ and $\sigma(q+1,k)$ is in $\sigma^g$ then the two operations are consecutive in $\sigma^g$ and, since the start time of the scheduled operations are fixed, $o(i,j)$ can be inserted between $\sigma(q,k)$ and $\sigma(q+1,k)$ at step $g$ which yields $t_{o(i,j)} = t'_{o(i,j)}$.
- If $\sigma(q,k)$ is in $\sigma^g$ and $\sigma(q+1,k)$ is not in $\sigma^g$, let $o1\neq\sigma(q+1,k)$ denote the operation consecutive to $\sigma(q,k)$ in $\sigma^g$. Since $\sigma(q+1,k)$ is inserted in $\sigma^g$ at a step strictly greater than $g$, we have $t_{\sigma(q+1,k)}+p_{\sigma(q+1,k)}+s_{\sigma(q+1,k),o1}<t_{o1}$. With (8) it comes: $\max\,(\,c_{o(i,j\text{-}1)}\,,\,c_{\sigma(q,k)}+s_{\sigma(q,k),o(i,j)})+p_{o(i,j)}+s_{o(i,j),\,\sigma(q+1,k)}\leq t_{o1}-s_{\sigma\,(q+1,k),o1}$

  Since the triangle inequality holds we have:

$$\max ( c_{o(i,j\text{-}1)} , c_{\sigma(q,k)}+s_{\sigma(q,k),o(i,j)}) + p_{o(i,j)} + s_{o(i,j),\, o1} \leq t_{o1}$$

which means that $o(i,j)$ could be inserted at step $g$ between $\sigma(q,k)$ and $o1$ which yields $t_{o(i,j)} = t'_{o(i,j)}$.

- If $\sigma(q,k)$ is not in $\sigma^g$ and $\sigma(q+1,k)$ is in $\sigma^g$, let $o2 \neq \sigma(q,k)$ denote the operation $\sigma(q+1,k)$ is consecutive to in $\sigma^g$. We can prove with similar arguments that $o(i,j)$ could be inserted at step $g$ between $o2$ and $\sigma(q+1,k)$ with $t_{o(i,j)} \leq t'_{o(i,j)}$.
- If neither $\sigma(q,k)$ nor $\sigma(q+1,k)$ are in $\sigma^g$, let $o3$ denote the operation scheduled on machine $k$ before time $t_{\sigma(q,k)}$ and let $o4$ denote the operation scheduled on machine $k$ after time $c_{\sigma(q+1,k)}$. We can prove with similar arguments that $o(i,j)$ could be inserted at step $g$ between $o3$ and $o4$ with $t_{o(i,j)} \leq t'_{o(i,j)}$.□

Note that the two strong assumptions of theorem 2 are the triangle inequality for setup times and the greedy behaviour of the insertion algorithm: the start time of an operation scheduled at a given step can not be modified during the remaining of the SGS. Since, the three proposed active SGSs belong to this category it follows from theorem 2 that any schedule generated by ExtendedGifflerThompsonSGS1, ExtendedGifflerThompsonSGS2 and SerialSGS is an active schedule.

To study the dominance properties of the SGS it remains to know whether any active schedule is obtainable by the proposed SGS. Let us illustrate the behaviour of the active algorithms with the example 1 presented in Section 1. We display in Figure 5 all the feasible active schedules[1]. The optimal solution is schedule (a) displayed on top of Figure 5.

********************* INSERT FIGURE 5 HERE *************************

In figures 6 and 7, we show the enumeration trees corresponding to all possible sets of eligible operations $E$ at each step of SerialSGS and ExtendedGifflerThompsonSGS1 or 2, respectively. Each level of the trees corresponds to a step of the algorithms. The arrow going down from a father node to a child node is issued from the operation selected for being scheduled at the step corresponding to the father node. A leaf of the tree corresponds to an active schedule.

********************* INSERT FIGURE 6 HERE *************************

[1] In this example, any semi-active schedule is also an active schedule

******************** INSERT FIGURE 7 HERE ************************

The comparison of the obtained trees brings the following remarks. First, the serial algorithm introduces a lot of symmetries. Indeed, solution (b) appears in 4 leaves of the tree. On the opposite the extended Giffler-Thompson algorithm does not produce symmetries since all the leaves of its tree correspond to distinct solutions. On the other hand, ExtendedGifflerThompsonSGS1 and 2 are unfortunately shown here to be non-dominant since the unique optimal solution (a) does not appear in the enumeration tree. The basic reason for the present example is that at step (1) operation $o(2,2)$ is not in the set $E$ of eligible operations because it is not available. However, because of the large setup time, $o(2,2)$ should be considered as conflicting with $o(1,1)$. Hence a direct extension of the Giffler-Thompson algorithm does not provide a set of dominant schedules. Theorem 3 shows that the serial SGS does not have the same drawback.

*Theorem* 3

The set of schedules generated by SerialSGS is dominant.

*Proof*

Let us consider a static priority vector $\pi$. Let $t$ denote the schedule generated by SerialSGS with $\pi$. Let $t'$ denote the schedule generated by SemiActiveSGS with $\pi$. SerialSGS and SemiActiveSGS only differ by the start time computation. Hence, the same operations are selected through $\pi$ by both algorithms at each step. For the selected operation $o(i^*,j^*)$, we have $t'_{o(i^*,j^*)} \geq t_{o(i^*,j^*)}$ if the triangle inequality (6) holds because in SerialSGS, $o(i^*,j^*)$ is possibly inserted in the sequence of its machine without delaying the already scheduled operations. Hence, the schedule obtained by SerialSGS with $\pi$ cannot be worse that the schedule obtained by SemiActiveSGS with $\pi$.

□

## 2.4. Non-delay SGSs

When setup times are not present, a schedule is said to be a non-delay schedule if no machine is left idle while it could start the processing of an operation, see Baker (1974). It follows that the classical non-delay schedule generation scheme, called the parallel SGS in resource-constrained project scheduling, see Kolisch (1996), does time incrementation.

Starting with time $\tau = 0$, at each iteration, an operation that can start at time $\tau$ is selected. Then $\tau$ is updated to the smallest possible start time of an unscheduled operation. Obviously, this definition does not suit the possible anticipation of setup times. Indeed, whenever an operation becomes available at a given time $\tau$ then in some cases it would have been possible to start its setup before time $\tau$.

However, we give two different definitions of a non-delay schedule in the SDST-JSP inspired by the earliest start policy of the non-delay concept.

*Definition* 3

A type 1 non-delay schedule is a feasible (active) schedule in which no machine is kept idle or performing a setup at any time while it could process an operation.

*Definition* 4

A type 2 non-delay schedule is a feasible (active) schedule in which no machine is kept idle at any time while it could process an operation or perform a setup in such a way that an operation can start immediately after the setup.

Roughly, type 1 non-delay schedules are “production-oriented” as they aim at reducing setups by prefering production to setup activities. Type 2 non-delay schedules are “time-oriented”. Note that if there are no setup times, both definitions are equivalent to the classical definition.

For the job-shop problem without setup times, it is well-known that the non-delay schedules are included in the set of active schedules but they are not dominant. Types 1 and 2 non-delay schedules have no chance to possess the dominance property since the JSP is a special case of the SDST-JSP. Obviously, they are also active schedules.

We describe an appending "non-delay" SGS extending the classical non-delay SGS for the JSP as described in Baker (1974) in the case a type 1 non-delay scheduled has to be generated.

- NonDelaySGS1 is an appending SGS

- the eligible set $E$ is computed as follows. Let $\tilde{o}$ denote the available operation of smallest start time (job index is used to break ties). We have:

(15) $$\tilde{o} = \text{argmin } \{\text{ESA}_o \mid o \in A\}$$

$E$ is the set of operations having the same start time as $\tilde{o}$, that is:

(16) $$E = \{o \mid o \in A,\ m_o = m_{\tilde{o}},\ \text{ESA}_{\tilde{o}} = \text{ESA}_o\ \}$$

When all setup times are equal to 0 the proposed algorithm behaves like the non-delay schedule generation scheme proposed in [4]. NonDelaySGS1 has an $O(NnT)$ time complexity.

*Theorem* 4

For any priority vector, NonDelaySGS1 generates a type 1 non-delay schedule and any type 1 non-delay schedule is generated by NonDelaySGS1 with a priority vector.

*Proof*

Suppose that $t$, a schedule generated by NonDelaySGS1 with a priority vector $\pi$, is not a type 1 non-delay schedule. It follows that we can find an idle time period $\tau$ and an operation $o(i,j)$ that can start at time $\tau$ while $\tau < t_{o(i,j)}$. Let $o1$ denote the first operation verifying $t_{o1}>\tau$. Let $g$ denote the step where $o1$ is scheduled by NonDelaySGS1. If $o(i,j)$ is available then, since we have $\text{ESA}_{o(i,j)} \leq \tau$ and $t_{o1}>\tau$, $o1$ should not appear in the eligible set, which is in contradiction with its definition. If $o(i,j)$ is not available let $o(i,j-x)$ with $x\geq1$, the first available operation of job $i$. We have $t_{o(i,j-x)} \leq \tau$ and $t_{o1}>\tau$. Hence $o1$ should not appear in the eligible set, a contradiction. It follows that any schedule generated by NonDelaySGS1 is a type 1 non-delay schedule. Let $t$ denote a type 1 non-delay schedule and let $\pi$ denote the priority vector obtained by sorting the operations in the non-decreasing order of the starting times in $t$. We see that $\pi$ is precedence-compatible, i.e. we have for any operation $o(i,j)$, $\pi_{o(i,j)} > \pi_{o(i,j-1)}$ whenever $p_{o(i,j-1)} > 0$. Hence at the step, $g$ the operation $o(i,j)$ such that $\pi_{o(i,j)} = g$ is in the eligible set and can be started at time $t_{o(i,j)}$.

□

We now study the insertion SGS extending the classical non-delay SGS for the JSP [4] proposed by Baker (1974) in the case a type 2 non-delay scheduled has to be generated.

- NonDelaySGS2 is an insertion SGS

- the eligible set $E$ is computed as follows. Let $\tilde{o}$ denote the available operation of smallest start time (job index is used to break ties). Recall that $\lambda(k)$ is the last operation scheduled on machine k. We have:

(17) $$\tilde{o} = \text{argmin}\,\{\text{ESA}_o - s_{\lambda(k),o} \mid o \in A\}$$

$E$ is the set of operations having the same setup start time as $\tilde{o}$, that is:

(18) $$E = \{o \mid o \in A,\ m_o = m_{\tilde{o}},\ \text{ESA}_{\tilde{o}} - s_{\lambda(k),\tilde{o}} = \text{ESA}_o - s_{\lambda(k),o}\}$$

When all setup times are equal to 0 the proposed algorithm behaves also like the non-delay schedule generation scheme proposed in [4]. NonDelaySGS2 has an $O(N(nT+m))$ time complexity.

We explain through example 4 why an insertion SGS is needed whereas the eligible set considers appending start times.

*Example* 4:
Consider a problem with 2 jobs and 4 machines with no setup times on machines 2, 3, 4 and non-zero setup times on machine 1. We have $m_{o(1,1)}=4$, $m_{o(1,2)}=1$, $m_{o(2,1)}=3$, $m_{o(2,2)}=2$, $m_{o(2,3)}=1$, $p_{o(1,1)}=6$, $p_{o(1,2)} = p_{o(2,1)}=2$, $p_{o(2,2)}=p_{o(2,3)}=1$. On machine 1, we have $s_{0,o(2,3)}=3$, $s_{0,o(1,2)}=5$ and $s_{o(2,3),\,o(1,2)}=2$. All other data are left undetermined. The triangle inequality holds. □

Suppose that, at steps 1 and 2 of the SGS, operations $o(1,1)$ and $o(2,1)$ are scheduled, respectively. Both operations being available at time 0, this is compatible with the conflict set definition 16. At step 3 of the algorithm available operations are $o(1,2)$, and $o(2,2)$. According to definition 15, operation $\tilde{o}$ is $o(1,2)$ with $\text{ESA}_{o(1,2)}-s_{0,o(1,2)}=1$ and the eligible set is $E=\{o(1,2)\}$ because $o(2,3)$ is not available. In figure 8, partial solution (a) corresponds to the selection of $o(1,2)$ at step 3.

********************* INSERT FIGURE 8 HERE ************************

Suppose now that $o(2,2)$ is scheduled at step 4. At step 4, $o(2,3)$ becomes available and can be inserted before $o(1,2)$ with $\text{ESI}_{o(2,3)}-s_{0,o(2,3)}=0$, as displayed in part (b). It follows that if an appending SGS is used, solution (a) does not yield a type 2 non-delay schedule.

Unfortunately, it has to be pointed out that the proposed non-delay SGS may generate a schedule out of the set of the type 2 non-delay schedules, even when used with an insertion scheme, as shown in example 5.

*Example* 5:

We modify example 4 simply by setting $s_{o(2,3),\, o(1,2)}$=3. □

In this case partial solution (b) of figure 8 cannot be generated when *o*(2,3) becomes available because *o*(2,3) cannot be inserted anymore before *o*(1,2) without increasing the start time of *o*(1,2) by 1. In this small example, the only type 2 non-delay schedule is unreachable by a SGS restricting the eligible set to available operations. This drawback is another illustration of the negative impact of sequence-dependent setup times on schedule properties.

In table 1, a summary of the theoretical results of our study of schedule generation schemes for the SDST-JSP is displayed, giving for each proposed SGS, the type of schedule it generates, its time complexity, its ability to generate all the schedules of its target type and the dominance property of the generated schedule set w.r.t. makespan minimization.

********************* INSERT TABLE 1 HERE ************************

## 3. Experimental comparison of the proposed SGSs

### 3.1. Benchmark instances

We have tested the proposed SGSs on SDST-JSP benchmark instances proposed by Brucker and Thiele (1996) (called BT instances in the sequel). These 15 instances are job-shop problems with setup times derived from the corresponding benchmark problems of Adams, Balas and Zawack (1988). In these instances, each operation has a setup type and all operations of the same job have the same setup type. Durations have been randomly generated between 1 and 100. The setup time matrix gives the time to switch from one setup type to another one on any machine. The problems are cast into 3 groups. The first group contains 5 problems of 5/10/5 machines/jobs/setup_types. The second group contains 5 problems of 5/15/5 machines/jobs/setup_types. The third group contains 5 problems of 5/20/10 machines/jobs/setup_types. The interest of using these instances is that optimal or good solutions have been generated either by the branch and bound of Brucker and Thiele (1996) or by the constraint programming technique of Focacci, Laborie and Nuijten (2000). It follows that the experiments will give an idea of the distance from optimum of the solution obtained by priority rule based heuristics. The setup times have been generated according to a quotient *q* of average setup time divided by the average processing time. Groups 1 and 2 have an average quotient $q \cong 0.5$ which corresponds to medium setup times while group 3 has an average quotient $q \cong 1$, giving larger setup times. However these setup times have a special structure. The setup time matrix for 5 setup types is the following for all instances:

```
10 20 30 40 50
0 10 20 30 40
40 0 10 20 30
30 40 0 10 20
20 30 40 0 10
10 20 30 40 0
```

The setup time matrix for 10 setup type is the following for all instances:

```
10 20 30 40 50 60 70 80 90 100
0 10 20 30 40 50 60 70 80 90
90 0 10 20 30 40 50 60 70 80
80 90 0 10 20 30 40 50 60 70
70 80 90 0 10 20 30 40 50 60
60 70 80 90 0 10 20 30 40 50
50 60 70 80 90 0 10 20 30 40
40 50 60 70 80 90 0 10 20 30
30 40 50 60 70 80 90 0 10 20
20 30 40 50 60 70 80 90 0 10
10 20 30 40 50 60 70 80 90 0
```

Hence it appears that this set of instances is rather small and too specific to make reliable comparisons. Hence we have generated other instances as follows. First, we have duplicated the 5/20/10 instances transforming them into 5/20/5 instances. Then, we have duplicated the 5/10/5 and 5/15/5 instances, transforming them into 5/10/10 and 5/15/10 instances. We obtain in this way 15 instances with 5 setup types and 15 instances with 10 setup types. Last, we have duplicated all 15 BT instances setting one distinct setup type per operation, following the general model presented in this paper. In this last set, each pair of distinct operations is associated with a random setup time between 0 and approximately 100. The triangle inequality is satisfied by associating random coordinates to operations. The setup time is then the Euclidean distance between the two points. We obtain in this way 15 instances with “more” random sequence-dependent setup times.

### 3.2. Priority rules

A priority rule $\pi$ gives a value to an operation $o(i,j)$. Depending on the rule, the operation selected in the set $E$ of candidate operations is the one having the smallest or the greatest value.

We have used the following priority rules, some being directly issued from classical JSP rules, others being adapted to deal with setup times.

- MWKR (Most WorK Remaining): $\pi_{o(i,j)} = \sum_{l=j}^{m} p_{o(i,l)}$
- SST (Shortest Setup Time): $\pi_{o(i,j)} = s_{\lambda(m_{o(i,j)}),o(i,j)}$
- SSTPT (Shortest Setup Time and Processing Time):

$\pi_{o(i,j)} = p_{o(i,j)} + s_{\lambda(m_{o(i,j)}),o(i,j)}$

- MINSLACK (Min Slack) $\pi_{o(i,j)} = \max_{l=j,\dots,m} (c_{o(i,l)})$
- MOPER (Most Operations Remaining) $\pi_{o(i,j)} = m+1-j$
- RAND (Random)
- MINSTART (Min Start time): $\pi_{o(i,j)} = ESA_{o(i,j)}$
- MINSTSTART (Min Setup Start time): $\pi_{o(i,j)} = ESA_{o(i,j)} - s_{\lambda(mo(i,j)),o(i,j)}$
- MINEND (Min End Time): $\pi_{o(i,j)} = ESA_{o(i,j)} + p_{o(i,j)}$

### 3.3. Single- and multi-pass heuristics

We have embedded the proposed SGSs into single- and multi-pass priority rule based heuristics. The single-pass heuristic consists in generating a unique solution with a given SGS associated with a given priority rule. For multi-pass methods, we have tested random sampling methods where, at each pass, the SGS is used with a random based priority rule. We have used a very simple random sampling procedure in which the priority rule is randomly biased for all SGSs using a random factor $\alpha$. More precisely, the operation of $E$ with the smallest priority value is selected unless another operation is selected with an $\alpha$ probability. Moreover, all ties are broken randomly instead of using the job index rule.

### 3.4. Computational results

*Comparison of the proposed SGSs within single- and multi-pass heuristics*

********************* INSERT TABLE 2 HERE *************************

********************* INSERT TABLE 3 HERE *************************

********************* INSERT TABLE 4 HERE *************************

Tables 2 to 4 give the results of the proposed SGSs on the 45 instances described in Section 3.1. Tables 2 gives the results of single-pass heuristics with all SGSs and all tested priority rules. Tables 3 and 4 give the results of multi-pass heuristics for 1000 and 10000 iterations, respectively. We give for each pair SGS/priority rule the number of problems out of 45 for which this combination found the best result (among all heuristics tested for the considered number of iterations) and, in brackets, the average deviation in percent from the best result obtained on each instance by the multi-pass heuristic with 10000 iterations. Hence, the former value gives an indication on the relative ranking of the pair SGS/rule for a given number of iterations whereas the latter value gives an absolute measure of performance w.r.t. the best known solution. In each table, row “BEST” gives the same indicators taking for each SGS and each instance the best result among all rules. ND, EGT, SA and SE stand for NonDelay, ExtendedGifflerThompson, SemiActive and Serial SGSs, respectively. We give in seconds the average and maximum CPU time per instance, on a Pentium 4 Personnal Computer with 2.40 GHz and 512 Mo RAM.

For the single pass heuristics (table 2), row “BEST” obtains the global ranking ND 1 > EGT1,EGT2 > SE,SA > ND 2. There is no significant difference between EGT 1 and 2. There is also no significant difference between SE and SA. This can be explained by examining the priority rule level. Except for ND 1, the best priority rule is MINSTART (Minimum appending Start time). Observe that when SA, SE, and EGT are applied in conjunction with the MINSTART priority rule, they all resort to ND 1. This can be verified in row MINSTART where all the results are identical to the ND 1 results. Indeed all these SGSs select for scheduling the operation with the minimum appending start time breaking the tie with the job index rule. The various priority rules used with ND 1 are just variants on the apparently superior MINSTART policy. ND 2 gives on the contrary the poorest results. Note that for ND 2, the SST and the MINSTART rule are equivalent, since ND 2 schedules operations of minimal setup start time. For all SGSs, the CPU times are negligible (less than 0.001 seconds).

Table 3 shows completely different results when the SGSs are applied in a multi-pass heuristic with 1000 iterations. The ranking of the SGSs when taking the best results among all rules becomes SE > SA > EGT 2> EGT 1 > ND 1 > ND 2.
ND 1 and ND 2 are significantly outperformed by the other SGSs since they are ranked only 2 or 3 times first and have an average deviation of about 8% from the best solution whereas all other SGSs are ranked from 16 to 26 times first and have an average deviation of about 2%. Globally, we see that applying a multi-start heuristic is fruitful, since the best average deviation from the best solution has been divided by more than 4 compared to table 2. The CPU times per instance are of less than 0.15 seconds for SA and SE. Looking at the priority rule level, the best rule is still MINSTART except for ND 1. Intuitively this result can bring to the idea that the MINSTART rule makes globally good decisions and that the best SGSs are the ones which are able to bring a sufficient level of diversification around this rule. Indeed, Random sampling with the ND 1 scheme only allows to select operations that have strictly the smallest earliest start time whereas with the serial scheme, the semi-active scheme and, more restrictively, the extended Giffler-Thompson scheme, an operation which does not start as soon as possible can be selected with the $\alpha$ probability, which brings more diversification.

The above hypothesis seems still valid for random sampling with 10000 iterations (table 4) since the SGSs ranking is now SE,SA < EGT1 < EGT2 < ND 1, ND 2. A surprising result is that there is no more a significant difference between the serial and the semi-active SGSs which means that when the number of iterations increases, the benefit of trying to keep the schedule active by insertion is lost. This allows to obtain the best results with an algorithm of low time requirements (less than 1.5 seconds per instance). Similarly EGT 1 is here slightly superior to EGT 2 whereas the latter SGS takes twice more time (more than 4 seconds per instance).

*Comparison of the proposed heuristics with truncated branch and bound*

******************** INSERT TABLE 5 HERE ***********************

Table 5 extracts from our testbed the 15 instances generated by Brucker and Thiele (1996) to evaluate the performance of the proposed SGSs in comparison with the state-of-the-art approaches. For each of the 15 instances and for each number of iterations, we display the makespan obtained by all SGSs and the average deviation from the previously best known results, obtained either by Brucker and Thiele (1996) or by Focacci, Laborie and Nuijten (2000). Note that the first 5 instances are the 5/10/5 instances, the next 5 instances are the 5/15/5 instances and the last 5 instances are the 5/20/10 instances. For the 5/10/5 instances, the optimum is known and displayed in column Cmax*. For the other instances, column Cmax* only displays an upper bound obtained by truncating the branch and bound of Brucker and Thiele after 2 hours, or the branch and bound of Foccacci *et al.* after 60 seconds. On one hand, the solution found by the SGSs are far from the optimum for the 5 small instances (from 2.51% to 18.84% for 10000 iterations). On the other hand, for the other instances, the proposed SGSs significantly improve the best known solutions in a very short time (less than 10 seconds per instance as shown in table 4). The improvement is from 2.08% to 10.86%. Even the single-pass heuristics improve very quickly (less than 0.001 seconds) the best known solutions on all 5/20/10 instances. For our best results (10000 iterations) we also display in brackets the average deviation over the lower bound computed by Brucker and Thiele (1996). The maximal deviation from this bound is of 9.89% on the 5/15/5 instances and of 19.28% on the 5/20/10 instances. Hence, the proposed heuristics give better results on medium and large instances than on smaller instances. They also outperform for these instances, the truncated branch and bound and the heuristics proposed by Brucker and Thiele (1996) run with a time limit of 2 hours on a SUN 4/20 workstation. We also obtain better results than the truncated branch and bound method of Focacci, Laborie and Nuijten (2000) run on a Pentium II 300 MHz with a time limit of 60 seconds. Looking closer to the priority-rule based heuristics proposed in Brucker and Thiele (1996), it appears that no MINSTART-based priority rule was used. Indeed, the two proposed SGSs are a parallel version of the semi-active SGS and an extended Giffler-Thompson SGS with conflict set definition (12), which does not take account of setup times. Hence exact methods would certainly benefit from integrating our heuristics.

**Commentaire [PL1] :** Je répète : n'est-ce pas 27.96 ?

*Influence of the number of setup types*

******************** INSERT TABLE 6 HERE ************************

********************* INSERT TABLE 7 HERE ************************

********************* INSERT TABLE 8 HERE ************************

We study the influence of the setup characteristics on the results. First, Tables 6, 7 and 8 display the results of the different SGSs for 5, 10 and *N* setup types instances, respectively. To simplify the reading, for each SGS and for each number of iterations, the best result among all rules is selected. We also display the best rule. Each table displays the results for 1, 1000 and 10000 iterations of the heuristic. Note that when the number of setup types is low, a lot of operations can be scheduled consecutively with zero setup time, which creates the possibility of grouping operations of the same type in order to save setups.

As a consequence of the preceding observation, for 1 and 1000 iterations, we observe that the performance of the best SGS decreases as the number of setup types increases. For 1000 and 10000 iterations, the gap between the best and the worst SGSs also increases as the number of setup types increases. These two observations underlines that the problem becomes more difficult as the number of setup types increases, i.e. when there is less possibility of grouping operations of the same setup type. Hence we conjecture that the new instances we propose are harder to solve by an exact method than the ones proposed by Brucker and Thiele (1996).

For any number of iterations, the performance of ND 1 strictly decreases as the number of setup types increases. ND 2 shows better performance for the large number of types than for the low number of types but it has its worse performance on the medium number of types. EGT 1 and 2 have their best performance for small and medium number of setup types and their worst performance for the large number of setup types. The performance of SA and SE also decreases as the number of setup types increases. Consequently, at this step of the analysis we can recommend to use the ND 1 SGS with a single-pass heuristic for any number of setup types (although the gap with the other SGSs is less significant for larger setup types). For multi-pass heuristics and small number of setup types, the serial SGS seems superior. For multi-pass heuristics with medium and large number of setup types, one can only discard the non-delay scheme, all other SGSs being able to provide good results.

*Influence of the setup time magnitude compared to processing times*

********************* INSERT TABLE 9 HERE ************************

********************* INSERT TABLE 10 HERE *************************

********************* INSERT TABLE 11 HERE *************************

********************* INSERT TABLE 12 HERE *************************

The influence of the magnitude of the setup times is now considered. We restrict to the set of 15 instances with *N* setup types for which we study 4 different magnitudes of setup times. Tables 9, 10, 11 and 12 display the results for zero, small, medium and large setup times, respectively. The processing times are all between 1 and 100. Small setup times are generated randomly between 0 and approximately 20. Medium setup times are generated between 0 and 100, which correspond to the value inside the 45 instances testbed. Large setup times are generated between 0 and 600.

For the instances with no setup time (Table 9), there is no more difference between type 1 and 2 non-delay SGSs. Similarly, EGT 1 and 2 resort to the Giffler-Thompson (GT) algorithm. The average deviation is given from the optimal solution known for the instances of Adams, Balas and Zawack (1988). With no setup times, the best results are obtained by the ND with the MOPER rule for 1 and 1000 iterations whereas GT obtains the best results with the MINSTART rule for 10000 iterations, being on average 1.33% above the optimal solution. In comparison, the results of the best heuristic on the 5 first BT instances with setup times for which the optimum is known (Table 5) are much worse: the smallest deviation from the optimum is of 2.51%. In accordance with the observations of Brucker and Thiele (1996), this shows that the presence of setup times increases considerably the difficulty of the problem.

For the single-pass heuristics with small and medium setup times (Tables 10 and 11), it appears that the non-delay SGSs are superior. ND 2 with the SST (and MINSTART) rule obtains the best results for small setup times whereas in accordance to the previous results, ND 1 obtains the best result (with the RAND rule) for medium setup times. For the other SGSs, the MINSTART rule is again ranked first. However for large setup times (table 12), the EGT 1 and 2 outperform the other SGSs. ND 1 is ranked last. For all SGSs except ND 1, the MINEND rule is ranked first, which underlines the impact of large setup times on priority rule performance.

For multi-pass heuristics, it seems that the magnitude of setup times has less influence, although the extended Giffler-Thompson SGS 1 and 2 obtain better results as the setup time magnitude increases. For multi-pass heuristics with 10000 iterations, the semi-active SGS is always better than the serial SGS and the gap increases as the setup time magnitude increases.

*Influence of the random factor $\alpha$*

********************* INSERT FIGURE 9 HERE *************************

We show in Figure 9 how the random factor $\alpha$ was selected. We show the average deviation from the best known solution on the 45 instances for the semi-active SGS with the MINSTART rule and 1000 iterations. It appears that satisfactory results are obtained with $\alpha < 1/6$. All the presented experimental results have been run with $\alpha=1/20$.

## 4. Concluding remarks

This paper gives a formal description of schedule sets and schedule generation schemes for the job-shop problem with sequence-dependent setup times. We have shown that some fundamental dominance properties are lost when considering simple extensions of the non-delay and Giffler-Thompson SGSs based on operation appending to sequence-dependent setup times, as performed by most previous studies on priority rule-based heuristics for the SDST-JSP in the literature, e.g. Allahverdi, Gupta and Aldowaisan (1999)[11] Kim and Bobrowski (1994), [14] and Ovacik and Uzsoy (1994). On the other hand, we have demonstrated that the serial algorithm based on operation insertion (Kolisch (1996)) has the ability to generate a dominant set of schedules.

In an experimental study we have tested priority-rule based heuristics on a set of standard benchmark instances and we have compared their results with the best solutions found by exact methods: Brucker and Thiele (1996), [7]. It appears that priority-rule based heuristics give medium quality results, even applied in multi-pass schemes with up to 10000 iterations, which underlines the difficulty of the problem and states that more sophisticated heuristics should be developed. However the proposed heuristics have found new upper bounds for 10 instances in a very short time, outperforming the truncated branch-and-bound algorithms proposed in Brucker and Thiele (1996) and [7]. Comparing the non-delay, Giffler-Thompson, semi-active and serial SGSs, we can give directions for the choice of the accurate SGS.

It appears that the serial SGS has an advantage — although it has never been mentioned in the above-mentionned studies — within multi-pass heuristics because of its theoretical ability to reach the optimum. This interest is confirmed experimentally because the serial SGS used with the MINSTART rule outperforms  the non-delay and extended Giffler-Thompson SGSs in multi-pass heuristics with 10000 iterations. However, when the number of setup types and when the magnitude of setup times is large, the semi-active SGS used with the MINSTART rule outperforms in turn the serial SGS while it requires a lower computational effort. The proposed extensions of the Giffler-Thompson algorithm obtain good results when the setup time magnitude is high. The proposed extensions of the non-delay SGS based on a MINSTART policy obtain the best results within single-pass heuristics and when the setup times is not too large. Non-delay SGSs obtain as a counterpart the worse results within multi-pass heuristics. This may be due to the fact that they bring too much reduction of the search space.

Further research may consist in adapting a setup based rule in conjunction with the MINSTART rule for the serial and semi-active SGSs. Other SGSs could also be developed. Following the work of Sotskov, Tautenhahn and Werner (1999) for sequence independent setup times it can be fruitful to search for other insertion based SGSs.

**Note**

This work has been supported by ECOS-CONICYT, project number C00E06.

*Table* 1. Summary of the theoretical results

| SGS | Schedule type | Time complexity | All | Dominant |
|---|---|---|---|---|
| SemiActiveSGS | Semi-Active | $O(NnT)$ | Yes | Yes |
| StrictOrderSGS | Active | $O(N(nT+m))$ | Yes | Yes |
| ExtendedGifflerThompsonSGS1 | Active | $O(N(nT+m))$ | No | No |
| ExtendedGifflerThompsonSGS2 | Active | $O(N(n^2T+m))$ | No | No |
| NonDelaySGS1 | Non Delay 1 | $O(NnT)$ | Yes | No |
| NonDelaySGS2 | Non Delay 2 and Active | $O(N(nT+m))$ | No | No |

*Table* 2. Results of single-pass heuristics on the 45 instances

| | ND 1 | ND 2 | EGT 1 | EGT 2 | SA | SE |
|---|---|---|---|---|---|---|
| SST | 12 (11.44) | 7 (13.56) | 4 (18.99) | 4 (18.90) | 0 (142.90) | 0 (38.99) |
| MOPER | 13 (11.97) | 0 (29.20) | 0 (30.79) | 0 (30.79) | 0 (33.55) | 0 (30.49) |
| SSTPT | **19 (11.37)** | 2 (25.09) | 0 (41.10) | 0 (41.10) | 0 (128.42) | 0 (45.23) |
| MINSLACK | 12 (11.59) | 0 (42.46) | 0 (52.89) | 0 (52.33) | 0 (366.06) | 0 (52.32) |
| MWKR | 12 (11.60) | 0 (35.28) | 0 (38.25) | 0 (38.23) | 0 (59.16) | 0 (40.22) |
| RAND | 14 (11.64) | 0 (40.45) | 0 (42.13) | 0 (42.34) | 0 (83.03) | 0 (47.40) |
| MINSTSTART | 12 (11.70) | 0 (38.64) | 0 (38.49) | 0 (38.49) | 0 (38.64) | 0 (38.64) |
| MINSTART | 12 (11.51) | 7 (13.56) | 12 (11.51) | 12 (11.51) | 12 (11.51) | 12 (11.51) |
| MINEND | **19 (11.46)** | 2 (25.09) | 2 (27.48) | 2 (27.48) | 2 (27.48) | 2 (27.48) |
| BEST | **30 (9.19)** | 9 (12.26) | 18 (10.17) | 18 (10.22) | 14 (11.07) | 14 (11.05) |
| AV CPU | < 0.001 | < 0.001 | < 0.001 | < 0.001 | < 0.001 | < 0.001 |
| MAX CPU | < 0.001 | < 0.001 | < 0.001 | < 0.001 | < 0.001 | < 0.001 |

*Table* 3. Results of multi-pass heuristics on the 45 instances (1000 iterations)

| | ND 1 | ND 2 | EGT 1 | EGT 2 | SA | SE |
|---|---|---|---|---|---|---|
| SST | 3 (7.92) | 2 (10.37) | 2 (5.88) | 3 (5.76) | 0 (44.10) | 0 (14.96) |
| MOPER | 2 (9.93) | 0 (18.64) | 0 (19.38) | 0 (19.12) | 0 (25.84) | 0 (21.82) |
| SSTPT | 1 (11.37) | 0 (23.75) | 0 (22.75) | 0 (22.75) | 0 (67.61) | 0 (25.35) |
| MINSLACK | 1 (11.59) | 0 (40.23) | 0 (38.45) | 0 (38.33) | 0 (237.91) | 0 (31.12) |
| MWKR | 1 (11.60) | 0 (34.78) | 0 (27.11) | 0 (27.07) | 0 (43.66) | 0 (26.94) |
| RAND | 3 (7.80) | 0 (20.35) | 0 (20.35) | 0 (20.05) | 0 (39.38) | 0 (23.13) |
| MINSTSTART | 2 (8.02) | 0 (22.61) | 0 (21.97) | 0 (21.24) | 0 (22.75) | 0 (22.31) |
| MINSTART | 3 (7.81) | 2 (10.37) | 14 (2.74) | 13 (2.55) | 17 (2.10) | **25 (1.98)** |
| MINEND | 1 (11.46) | 0 (23.75) | 0 (14.79) | 1 (14.86) | 0 (13.36) | 0 (13.95) |
| BEST | 3 (7.79) | 2 (8.00) | 16 (2.21) | 17 (2.06) | 17 (1.98) | **26 (1.86)** |
| AV CPU | 0.22 | 0.28 | 0.25 | 0.42 | 0.12 | 0.15 |
| MAX CPU | 0.47 | 0.64 | 0.52 | 0.92 | 0.28 | 0.34 |

*Table* 4. Results of multi-pass heuristics on the 45 instances (10000 iterations)

| | ND 1 | ND 2 | EGT 1 | EGT 2 | SA | SE |
|---|---|---|---|---|---|---|
| SST | 1 (7,92) | 2 (10.37) | 4 (3.70) | 2 (4.13) | 0 (36.19) | 0 (11.48) |
| MOPER | 1 (9,93) | 0 (18.42) | 0 (16.73) | 0 (16.96) | 0 (24.16) | 0 (20.21) |
| SSTPT | 0 (11.37) | 0 (23.75) | 0 (20.25) | 0 (20.23) | 0 (58.41) | 0 (21.93) |
| MINSLACK | 0 (11.59) | 0 (40.21) | 0 (35.35) | 0 (35.32) | 0 (212.28) | 0 (26.81) |
| MWKR | 0 (11.60) | 0 (34.78) | 0 (25.10) | 0 (24.71) | 0 (40.01) | 0 (24.52) |
| RAND | 1 (7.76) | 0 (17.30) | 0 (17.18) | 0 (16.64) | 0 (33.60) | 0 (19.60) |
| MINSTSTART | 1 (8.02) | 0 (22.73) | 1 (18.53) | 0 (18.89) | 0 (19.68) | 0 (19.42) |
| MINSTART | 1 (7.80) | 2 (10.37) | 13 (1.34) | 14 (1.32) | **23 (0.65)** | 20 (0.73) |
| MINEND | 0 (11.46) | 0 (23.75) | 1 (12.65) | 1 (12.77) | 0 (11.48) | 2 (11.25) |
| BEST | 1 (7.76) | 2 (7.42) | 18 (0.81) | 16 (1.04) | **23 (0.64)** | **22 (0.63)** |
| AV CPU | 2.07 | 2.81 | 2.49 | 4.23 | 1.24 | 1.50 |
| MAX CPU | 4.41 | 5.75 | 5.13 | 9.30 | 2.72 | 3.38 |

*Table* 5. Best results of single and multi-pass heuristics on the 15 BT instances

| It | 1 | | 1000 | | 10000 | | Cmax* |
|---|---|---|---|---|---|---|---|
| BT instance | Cmax | ΔCmax* | Cmax | ΔCmax* | Cmax | ΔCmax* (ΔLB) | |
| 1 | 844 | 5,76 | 826 | 3,51 | 818 | 2,51 | **798** |
| 2 | 959 | 22,32 | 869 | 10,84 | 829 | 5,74 | **784** |
| 3 | 842 | 27,96 | 803 | 22,04 | 782 | 18,84 | **658** |
| 4 | 789 | 25,84 | 745 | 18,82 | 745 | 18,82 | **627** |
| 5 | 733 | 12,25 | 704 | 7,81 | 704 | 7,81 | **653** |
| 6 | 1057 | 0,09 | 1026 | -2,84 | **1026** | -2,84 (4.06) | 1056 |
| 7 | 1104 | 1,56 | 1044 | -3,96 | **1033** | -4,97 (9.89) | 1087 |
| 8 | 1052 | -4,01 | 1034 | -5,66 | **1002** | -8,58 (9.75) | 1096 |
| 9 | 1101 | -1,61 | 1061 | -5,18 | **1060** | -5,27 (5.89) | 1119 |
| 10 | 1150 | 8,70 | 1052 | -0,57 | **1036** | -2,08 (2.78) | 1058 |
| 11 | 1617 | -2,47 | 1536 | -7,36 | **1478** | -10,86 (11.80) | 1658 |
| 12 | 1424 | -1,66 | 1333 | -7,94 | **1319** | -8,91 (15.80) | 1448 |
| 13 | 1463 | -5,55 | 1439 | -7,10 | **1439** | -7,10 (15.12) | 1549 |
| 14 | 1499 | -5,84 | 1492 | -6,28 | **1492** | -6,28 (6.42) | 1592 |
| 15 | 1684 | -3,44 | 1575 | -9,69 | **1559** | -10,61 (19.28) | 1744 |

*Table* 6. Results of the SGSs (1, 1000 and 10000 it) on the 15 instances with 5 setup types

| | ND 1 | ND 2 | EGT 1 | EGT 2 | SA | SE |
|---|---|---|---|---|---|---|
| 1 | **13 (7.57)** **MINEND** | 1 (10.89) SST | 4 (9.24) MINSTART | 4 (9.24) MINSTART | 4 (9.67) MINSTART | 4 (9.67) MINSTART |
| 1000 | 2 (4.99) MINSTSTART | 1 (4.48) SST | 6 (1.82) MINSTART | 4 (1.82) MINSTART | **5 (1.51)** **MINSTART** | **8 (1.67)** **MINSTART** |
| 10000 | 1 (4.92) RAND | 1 (3.68) SST | 5 (0.95) MINSTART | 7 (0.82) MINSTART | 6 (0.42) MINSTART | **11 (0.39)** **MINSTART** |

*Table* 7. Results of the SGSs (1, 1000 and 10000 it) on the 15 instances with 10 setup types

| | ND 1 | ND 2 | EGT 1 | EGT 2 | SA | SE |
|---|---|---|---|---|---|---|
| 1 | **9 (8.68)** **SSTPT** | 2 (13.14) SST | 9 (8.74) MINSTART | 9 (8.74) MINSTART | 5 (10.76) MINSTART | 5 (10.76) MINSTART |
| 1000 | 1 (7.81) RAND | 1 (10.86) SST | 8 (1.71) MINSTART | **9 (1.58)** **MINSTART** | 6 (1.96) MINSTART | **9 (1.74)** **MINSTART** |
| 10000 | 0 (7.81) RAND | 1 (10.28) SST | **10 (0.32)** **MINSTART** | 5 (1.03) MINSTART | **10 (0.54)** **MINSTART** | 6 (0.67) MINSTART |

*Table* 8. Results of the SGSs (1, 1000 and 10000 it) on the 15 instances with *N* setup types

| | ND 1 | ND 2 | EGT 1 | EGT 2 | SA | SE |
|---|---|---|---|---|---|---|
| 1 | **8 (11.32)** **RAND** | 6 (12.76) SST | 5 (12.53) MINSTART | 5 (12.69) MINSTART | 5 (12.68) MINSTART | 5 (12.71) MINSTART |
| 1000 | 0 (10.56) RAND | 0 (8.66) SST | 1 (3.40) MINSTART | 4 (2.79) MINSTART | 6 (2.49) MINSTART | **9 (2.18)** **MINSTART** |
| 10000 | 0 (10.56) RAND | 0 (8.31) RAND | 3 (1.15) MINSTART | 4 (1.26) MINSTART | **7 (0.95)** **MINSTART** | **5 (0.84)** **MINSTART** |

*Table* 9. Results of the SGSs (1, 1000 and 10000 it) on the 15 instances with no setup time

| | ND | GT | SA | SE |
|---|---|---|---|---|
| 1 | **15 (5.41*) MOPER** | 5 (8.56*) MWKR | 2 (12.62*) MOPER | 5 (9.15*) MWKR |
| 1000 | **13 (1.62*) MOPER** | 12 (1.64*) MOPER | 9 (2.15*) MINSTART | 12 (1.91*) MINSTART |
| 10000 | 12 (1.48*) RAND | **14 (1.33*) MINSTART** | 11 (1.88*) MINSTART | 11 (1.73*) MINSTART |

*Table* 10. Results of the SGSs (1, 1000 and 10000 it) on the 15 instances with small setup times

| | ND 1 | ND 2 | EGT 1 | EGT 2 | SA | SE |
|---|---|---|---|---|---|---|
| 1 | **7 (10.74) MINEND** | **8 (9.18) MINSTART** | 4 (11.36) MINSTART | 4 (11.36) MINSTART | 2 (12.96) MINSTART | 3 (12.23) MINSTART |
| 1000 | 0 (9.06) RAND | 1 (4.38) RAND | 0 (2.52) MINSTART | 2 (2.19) MINSTART | 4 (1.72) MINSTART | **11 (1.54) MINSTART** |
| 10000 | 0 (9.06) RAND | 1 (3.66) RAND | 1 (1.01) MINSTSTART | 2 (0.69) MINSTART | **9 (0.26) MINSTART** | 6 (0.35) MINSTART |

*Table* 11. Results of the SGSs (1, 1000 and 10000 it) on the 15 instances with medium setup times

| | ND 1 | ND 2 | EGT 1 | EGT 2 | SA | SE |
|---|---|---|---|---|---|---|
| 1 | **8 (11.32) RAND** | 6 (12.76) SST | 5 (12.53) MINSTART | 5 (12.69) MINSTART | 5 (12.68) MINSTART | 5 (12.71) MINSTART |
| 1000 | 0 (10.56) RAND | 0 (8.66) SST | 1 (3.40) MINSTART | 4 (2.79) MINSTART | 6 (2.49) MINSTART | **9 (2.18) MINSTART** |
| 10000 | 0 (10.56) RAND | 0 (8.31) RAND | 3 (1.15) MINSTART | 4 (1.26) MINSTART | **7 (0.95)** MINSTART | **5 (0.84)** MINSTART |

*Table* 12. Results of the SGSs (1, 1000 and 10000 it) on the 15 instances with large setup times

| | ND 1 | ND 2 | EGT 1 | EGT 2 | SA | SE |
|---|---|---|---|---|---|---|
| 1 | 4 (11.47) MINSLACK | 6 (10.57) MINEND | **8 (9.94) MINEND** | **8 (9.94) MINEND** | 6 (10.13) MINEND | 6 (10.13) MINEND |
| 1000 | 0 (11.47) RAND | 0 (8.82) MINSTART | 4 (2.01) MINSTART | **5 (1.98) MINSETUP** | 3 (1.78) MINSTART | **5 (1.88) MINSTART** |
| 10000 | 0 (11.47) RAND | 0 (8.42) MINSTART | 3 (0.74) MINSTART | 4 (0.91) MINSETUP | **10 (0.35) MINSTART** | 3 (0.77) MINSTART |

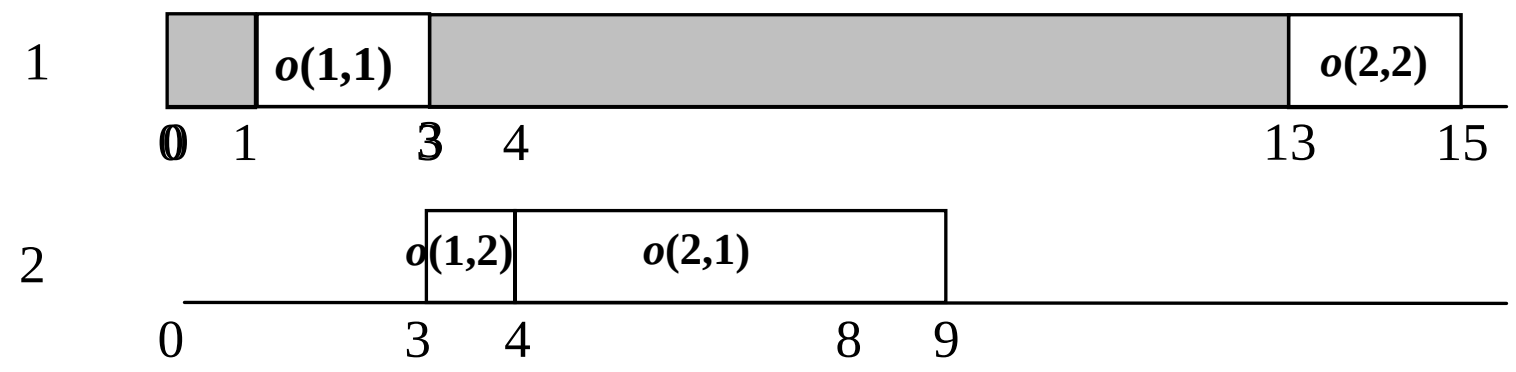

*Figure* 1. A feasible solution for example 1

**Algorithm** SGS($\pi$)

Initialize the available set $A$ with operations $o(i,1)$, $i = 1,\ldots,n$

**For** $q = 1,\ldots,N$ **do**

  Compute eligible set $E \subseteq A$ (depends on schedule category)

  Select operation $o(i^*,j^*) = \text{argmin}\{\pi_{o(i,j)} \mid o(i,j) \in E\}$

  Set $t_{o(i^*,j^*)} = \text{ESA}_{o(i^*,j^*)}$ ($t_{o(i^*,j^*)} = \text{ESI}_{o(i^*,j^*)}$) for an appending (insertion) SGS

  Set $A = A \setminus \{o(i^*,j^*)\}$ ($\cup o(i^*,j^*+1)$ if $j^* < m$)

**End For**

**return** $t$

**End** SGS

*Figure* 2. General framework of a SGS

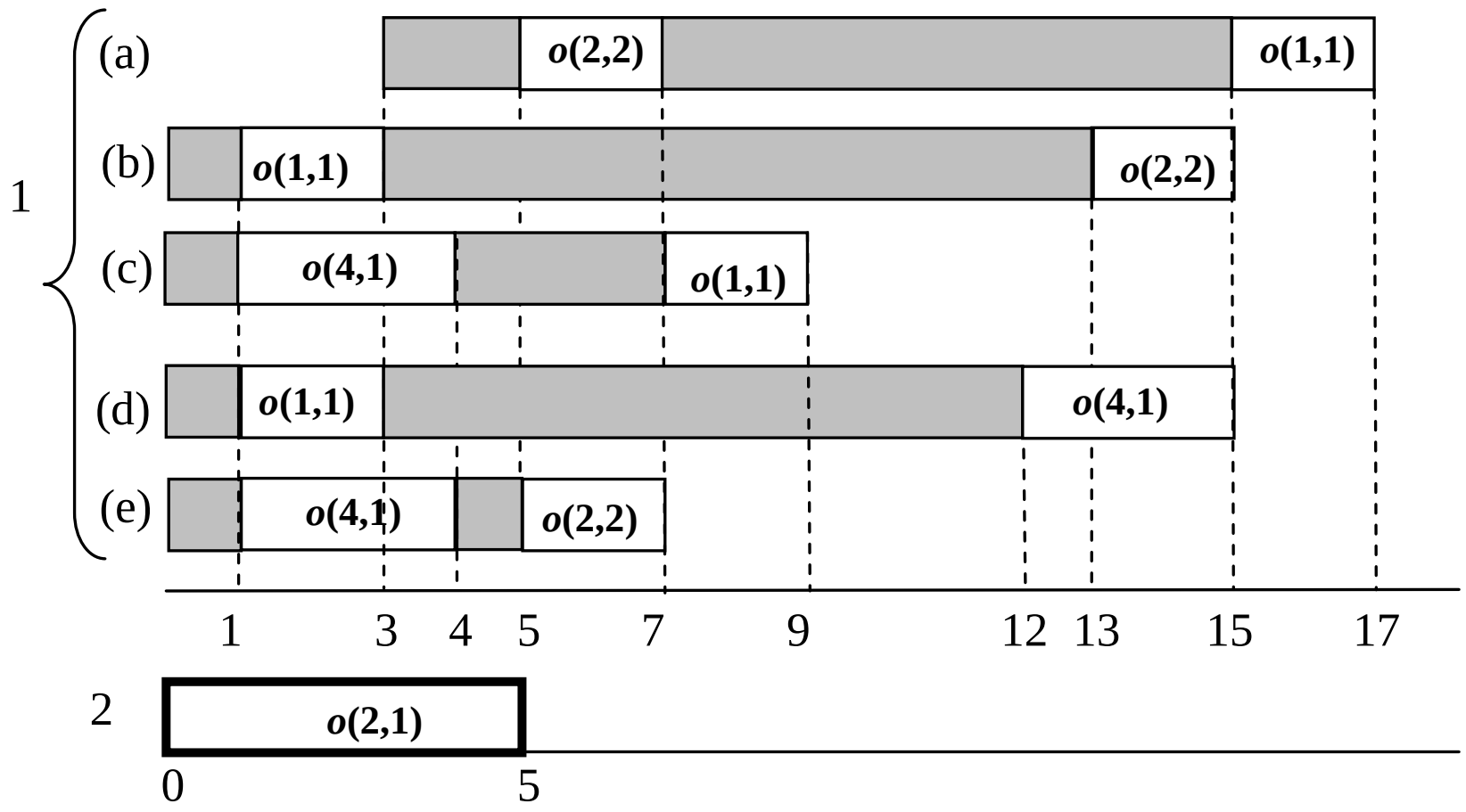

*Figure* 3. Partial active and non active solutions for Example 2

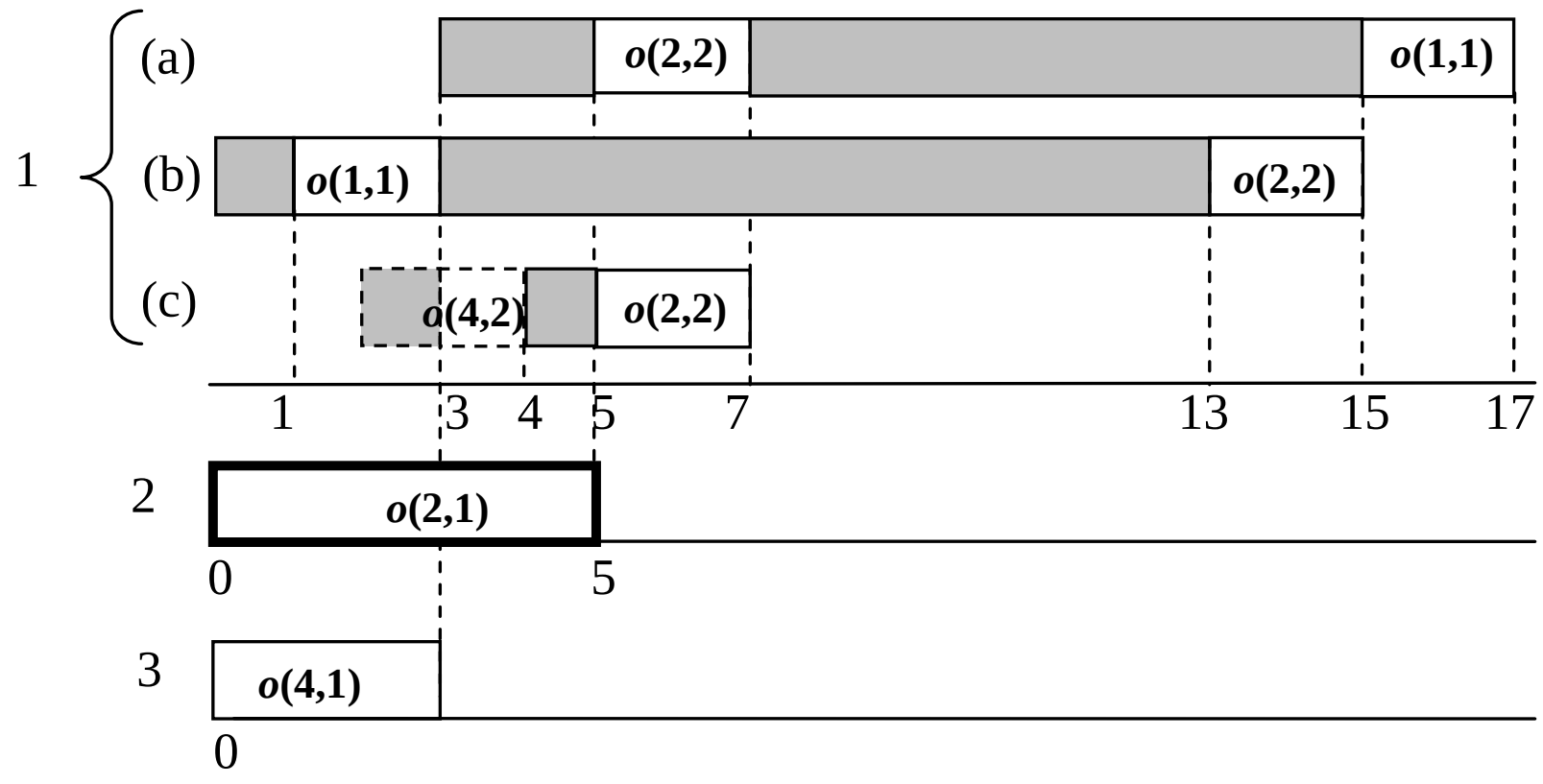

*Figure* 4. Partial active and non active solutions for Example 3

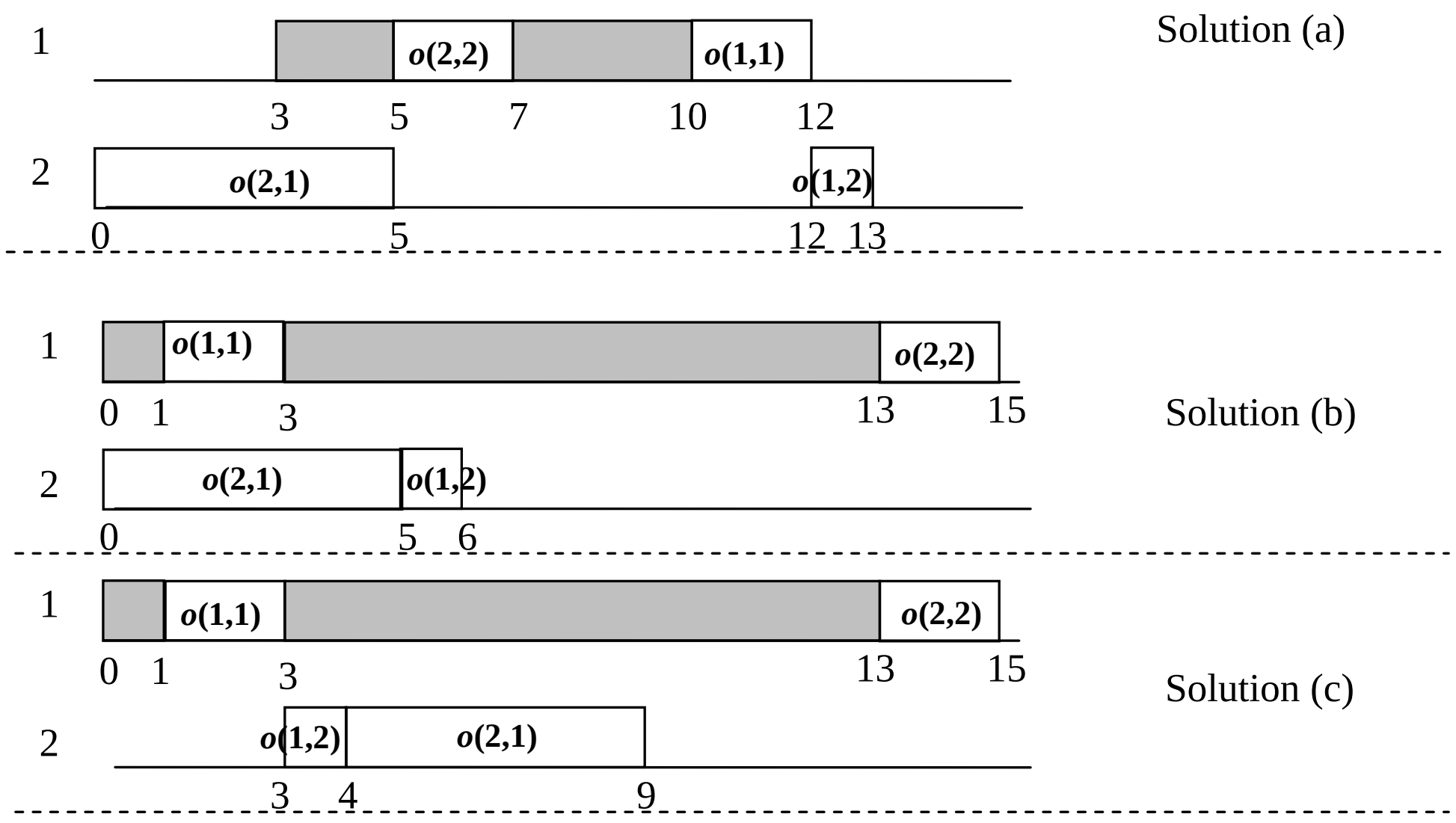

*Figure 5.* The active schedules of example 1

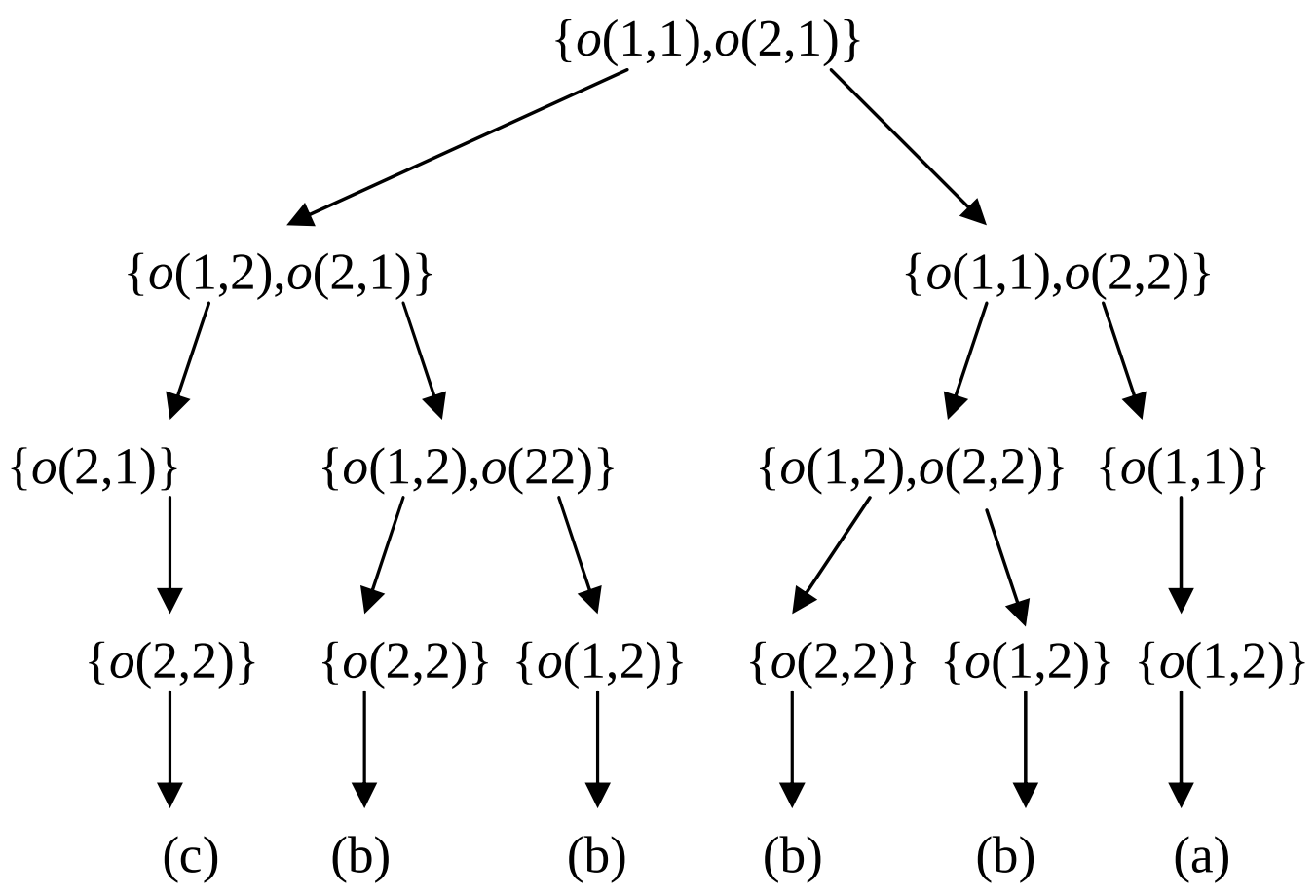

*Figure 6.* The Serial SGS enumeration tree

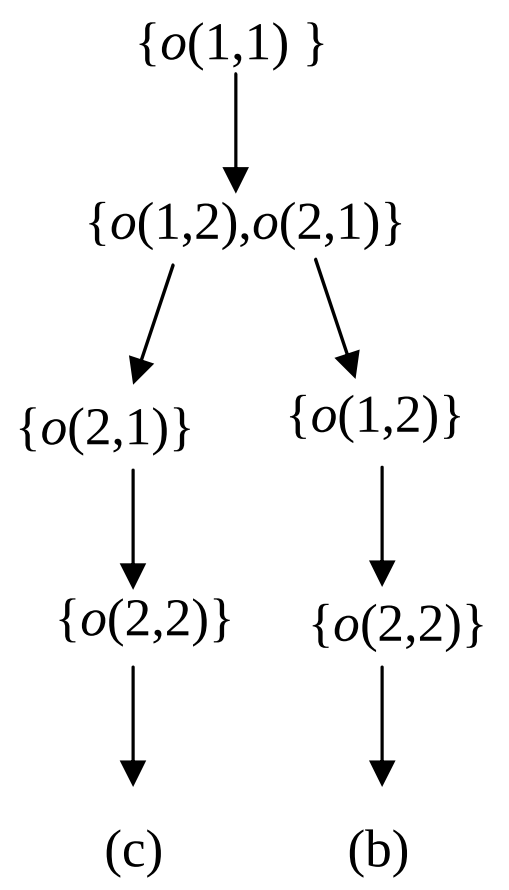

*Figure* 7 . The Extended Giffler-Thompson SGS1 and 2 enumeration tree

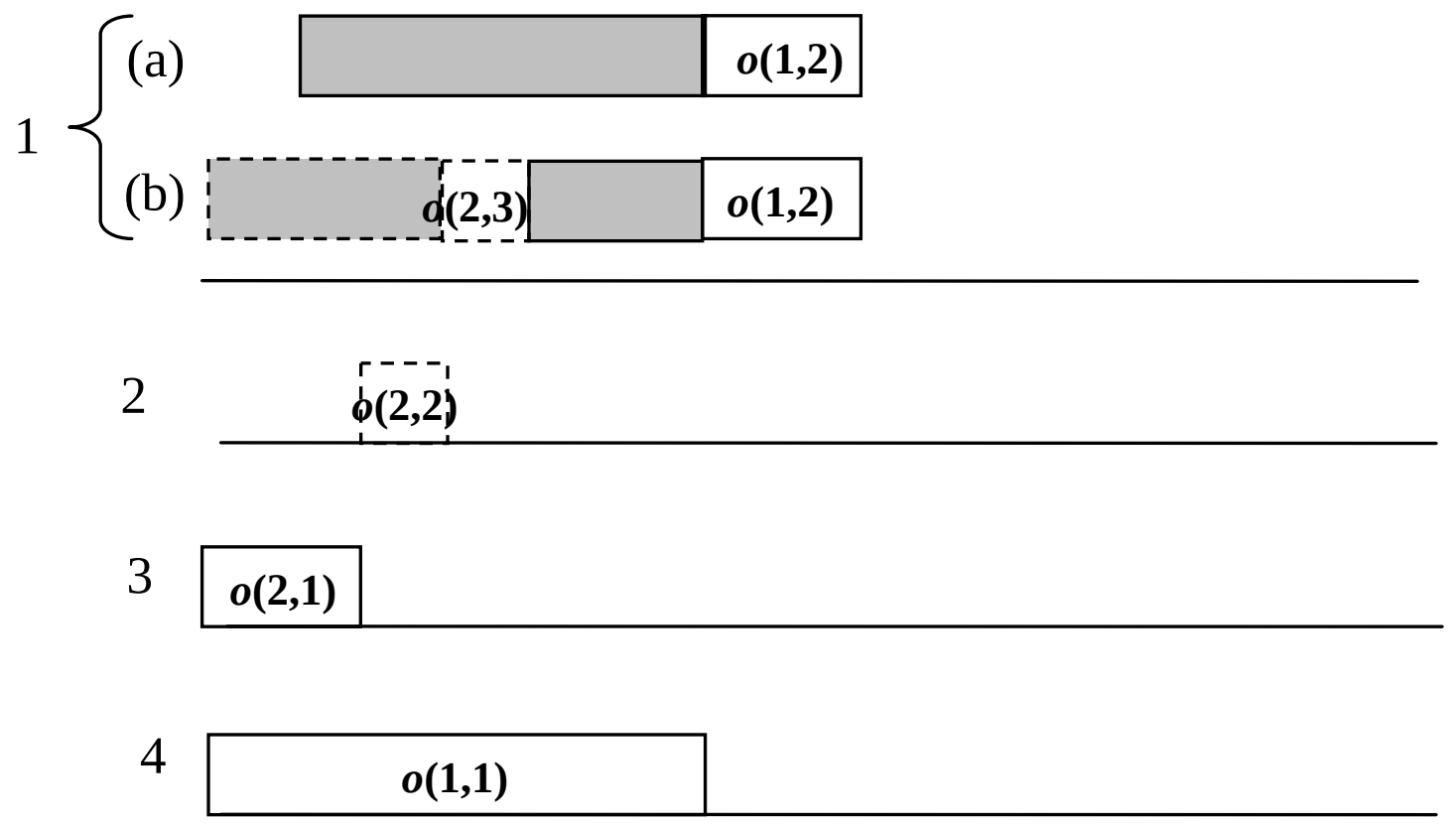

*Figure* 8. Partial solutions for example 4 in and out of the type 2 non-delay set

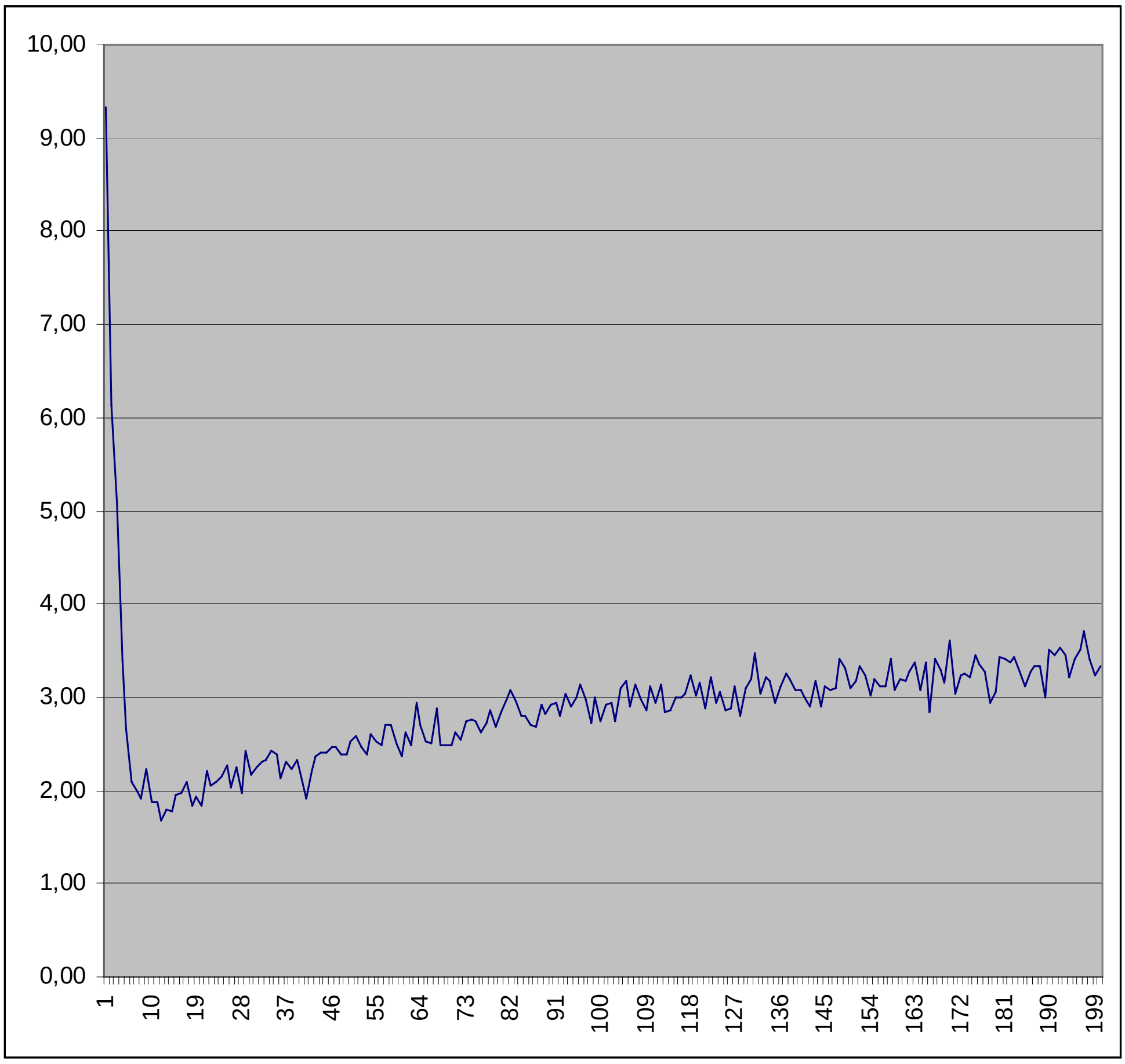

*Figure* 9. Quality of the solution of Semi-Active SGS with MINSTART for different random factors 1/α